\documentclass[fleqn,usenatbib]{mnras}

\usepackage{newtxtext,newtxmath}
\usepackage[T1]{fontenc}

\DeclareRobustCommand{\VAN}[3]{#2}
\let\VANthebibliography\thebibliography
\def\thebibliography{\DeclareRobustCommand{\VAN}[3]{##3}\VANthebibliography}

\usepackage{graphicx}  % Including figure files
\usepackage{amsmath}   % Advanced maths commands

\usepackage[separate-uncertainty=true, group-digits=integer]{siunitx}  % Provides units and their spacing
\usepackage{physics}   % Additional symbols like derivatives, erf etc.
\usepackage{subcaption}  % Multiple graphics in one figure
\usepackage{pdflscape}	% Landscape pages
\usepackage{xcolor}  % Get a nice color for edits
\definecolor{darkpastelgreen}{rgb}{0.01, 0.75, 0.24}
\newcommand{\ed}[1]{#1}

\newcommand{\DM}{\ensuremath{\mathrm{DM}}}

\DeclareSIUnit \jansky {Jy}
\DeclareSIUnit \parsec {pc}
\DeclareSIUnit \dmu {\parsec\per\centi\meter\cubed}

\graphicspath{{./}{figures/}}
\title[The FRB~20121102A November rain]{The FRB~20121102A November rain in 2018 observed with the Arecibo Telescope}

\author[J. N. Jahns et al.]{J. N. Jahns,$^{1}$\thanks{E-mail: \href{mailto:jjahns@mpifr-bonn.mpg.de}{jjahns@mpifr-bonn.mpg.de}}
	L. G. Spitler,$^{1}$
	K. Nimmo,$^{2,3}$
	D. M. Hewitt,$^{3}$
	M. P. Snelders,$^{3}$
	A. Seymour,$^{4}$
	\newauthor
	J. W. T. Hessels,$^{3,2}$
	K. Gourdji,$^{3,5}$
	D. Michilli,$^{6,7}$
	G. H. Hilmarsson$^{1}$
	\\
	$^{1}$Max-Planck-Institut für Radioastronomie, Auf dem Hügel 69, D-53121 Bonn, Germany\\
	$^{2}$ASTRON, Netherlands Institute for Radio Astronomy, Oude Hoogeveensedijk 4, 7991 PD Dwingeloo, The Netherlands\\
	$^{3}$Anton Pannekoek Institute for Astronomy, University of Amsterdam, Science Park 904, 1098 XH Amsterdam, The Netherlands\\
	$^{4}$Green Bank Observatory, P.O. Box 2, WV 24944, Green Bank, USA \\
	$^{5}$Centre for Astrophysics and Supercomputing, Swinburne University of Technology, Hawthorn VIC 3122, Australia\\
	$^{6}$MIT Kavli Institute for Astrophysics and Space Research, Massachusetts Institute of Technology, 77 Massachusetts Ave, Cambridge, MA 02139, USA\\
	$^{7}$Department of Physics, Massachusetts Institute of Technology, 77 Massachusetts Ave, Cambridge, MA 02139, USA
}

\date{Accepted XXX. Received YYY; in original form ZZZ}

\pubyear{2022}

\begin{document}
\label{firstpage}
\pagerange{\pageref{firstpage}--\pageref{lastpage}}
\maketitle

\begin{abstract}
We present 849 new bursts from FRB~20121102A detected with the 305-m Arecibo Telescope. Observations were conducted as part of our regular campaign to monitor activity and evolution of burst properties.
The 10 reported observations were carried out between \num{1150} and \SI{1730}{\MHz} and fall in the active period around November 2018. All bursts were dedispersed at the same dispersion measure and are consistent with a single value of \SI{562.4(1)}{\dmu}.
The rate varies between 0 bursts and \num{218(16)} bursts per hour, the highest rate observed to date.
The times between consecutive bursts show a bimodal distribution. 
We find that a Poisson process with varying rate best describes arrival times with separations $>\SI{0.1}{\s}$. Clustering on timescales of \SI{22}{\ms} reflects a characteristic timescale of the source and possibly the emission mechanism.
We analyse the spectro-temporal structure of the bursts by fitting 2D Gaussians with a temporal drift to each sub-burst in the dynamic spectra. We find a linear relationship between the sub-burst's drift and its duration. At the same time, the drifts are consistent with coming from the sad-trombone effect. This has not been predicted by current models.
The energy distribution shows an excess of high energy bursts and is insufficiently modelled by a single power-law even within single observations. 
We find long-term changes in the energy distribution, the average spectrum, \ed{and the sad-trombone drift,} compared to earlier and later published observations.
Despite the large burst rate, we find no strict short-term periodicity.
\end{abstract}

\begin{keywords}
	fast radio bursts -- methods: observational -- methods: data analysis -- radio continuum: transients
\end{keywords}

%%%%%%%%%%%%%%%%%%%%%%%%%%%%%%%%%%%%%%%%%%%%%%%%%%

%%%%%%%%%%%%%%%%% BODY OF PAPER %%%%%%%%%%%%%%%%%%

\section{Introduction}
\label{sec:intro}

Fast radio bursts (FRBs) are millisecond-duration flashes of radio waves that are of extragalactic origin \citep{Lorimer2007}. The sources of FRBs and their emission mechanisms are still uncertain \citep[see][for recent reviews]{Petroff2021,Lyubarsky2021}, but growing numbers of FRBs \citep{CHIME2021} and localizations within host galaxies \citep[e.g.][]{Bannister2019, Bhandari2021} recently boosted our understanding of their statistical properties. 

Observationally, FRBs are divided into repeaters and FRBs that have so far only been detected once. \citet{CHIME2021} found that these two populations indeed have different statistical properties, with repeaters having narrower bandwidths and longer durations on average \citep{Pleunis2021}. Moreover, the tendency of sub-bursts to drift to lower frequencies within a burst \citep{Hessels2019} -- now called \textit{sad-trombone effect} -- is characteristic for repeaters, while being very rare in (apparent) non-repeaters. This suggests that the presence of two distinct populations is due to different underlying source classes or emission mechanisms, even if it turns out that all FRBs will repeat eventually.

FRB~20121102A (hereafter FRB~121102) is the first discovered \citep{Spitler2014} and most studied repeater. It was observed to repeat by \citet{Spitler2016}, which allowed interferometric follow-up observations that localized it to a low-metalicity, star-forming dwarf galaxy at redshift $z=0.193$ \citep{Chatterjee2017, Tendulkar2017, Bassa2017}. The source is co-located with a compact, persistent radio source within \SI{40}{\parsec} projected distance \citep{Marcote2017}. Observations between 4--$\SI{8}{\GHz}$ revealed nearly \SI{100}{\percent} linear polarization with an extremely high rotation measure of $\sim$\SI{e5}{\radian\per\square\meter} \citep{Michilli2018, Gajjar2018}. The rotation measure decreases by an average of \SI{15}{\percent} per year, with percent level fluctuations between weeks \citep{Hilmarsson2021}. Together these findings are interpreted as the FRB being in the magneto-ionic environment of a massive black hole, a supernova remnant, or perhaps the wind nebula (plerion) of a young neutron star\ed{; the observations are also consistent with a binary model, if there is a massive stellar companion \citep[e.g.][]{Tendulkar2021}}. Several years of observations with different telescopes have also revealed periodic activity with a period of \SI{\sim160}{days} and a duty cycle of about \SI{60}{\percent} \citep{Rajwade2020,Cruces2021}. Reasons for this could either be that the source is in a binary system with a corresponding period \citep[e.g.][]{Gu2020,Lyutikov2020,Ioka2020,Kuerban2021,Du2021,Wada2021}, it is precessing \citep[e.g.][]{Levin2020,Sobyanin2020,Zanazzi2020,Sridhar2021}, or that it has a thus far unobserved slow period \citep{Beniamini2020}.

To further understand the source and its emission mechanism, it is crucial to study the burst properties and statistics of large burst samples. Such studies have shown clear spectro-temporal structures in the bursts of FRB~121102 as well as the sad-trombone effect \citep[see e.g.{}][]{Hessels2019}. Searches have been most fruitful at \SI{1.4}{\GHz}, where the source shows high activity. The largest numbers of bursts were found with the Arecibo Telescope and the Five-hundred-meter Aperture Spherical radio Telescope (FAST). \citet{Hewitt2021} reported a total of 478 bursts in \SI{59}{hours} of observations with the Arecibo Telescope through 2016 and an activity peak in September (a subset of which was previously found by \citet{Gourdji2019} and \citet{Aggarwal2021}). \citet{Li2021} reported 1652 bursts in 59.5 observing hours with FAST in September and October 2019, triggered when the source was known to be active. Further studies were conducted with the Effelsberg telescope 
\citep{Hardy2017, Houben2019, Cruces2021}, Apertif \citep{Oostrum2020}, the Lovell telescope \citep{Rajwade2020}, as well as the MeerKAT and Nan\c{c}ay telescopes \citep{Platts2019,Caleb2020}.

From the properties that a large burst sample permits to study, the energy distribution, the wait-times, and the spectro-temporal structure are among the most interesting.
The energy distribution can give important insight into the emission mechanism and allows direct comparison to rotation-powered pulsars and magnetically powered magnetars, which are well-studied candidate progenitor classes. Normal pulses of some pulsars form a log-normal energy distribution \citep[see e.g.{}][]{Burke-Spolaor2012}, whereas energies of giant pulses follow a power-law \citep[see e.g.][]{Bera2019, Abbate2020}. Similarly, burst energies of magnetars also show log-normal distributions with a tail at high energies \citep{Lynch2015} and power-laws in X-rays \citep{Gogus1999, Gogus2000}. The bursts of FRB~121102 were found to have different \ed{power-law indices in different studies;} for example \citet{Gourdji2019} reported $\gamma=-1.8$, while \citet{Cruces2021} found $\gamma=-1.1$. \citet{Li2021} found a bimodal energy distribution, which notably evolved during their observing campaign. \citet{Hewitt2021} reported a separation of bursts into two groups in the parameter space of energy, width, and bandwidth. It now becomes evident that most of these differences come from actual temporal changes in the emission energies and properties, rather than selection effects.

Analysis of the wait-times (the time between consecutive \ed{\mbox{(sub-)}}bursts) can reveal characteristic timescales or changes in the activity of a repeater. The discovery of the \SI{\sim160}{days} periodicity has explained some of the clustering previously observed \citep{Oppermann2018}, yet changes in the rate on timescales of hours to days still persist. Furthermore, short wait-times \citep[first discussed in][]{Katz2018, Zhang2018a, Gourdji2019} have appeared in several studies of high-rate observations as a second peak in the wait-time distribution \citep{Li2019,Li2021, Aggarwal2021, Hewitt2021}. \citet{Cruces2021} have shown that bursts within single observations still follow Poisson statistics if one excludes wait-times of $<\SI{1}{\s}$. \ed{However, the rate is not constant and has been seen to change on timescales of minutes at \SI{6}{\GHz} \citep{Gajjar2018}.}

Apart from the sad-trombone effect, a second more subtle, possibly related spectro-temporal effect has been reported by \citet{Rajabi2020} and \citet{Chamma2021}. Here, within one sub-burst lower frequencies arrive later than higher frequencies depending on the burst width and frequency. This \textit{intra-burst drift} is so far only explained by a family of models where the FRB's emission region moves with relativistic speed and where the emission process has a time delay between a trigger and the emission. It was particularly discussed in the framework of Dicke's superradiance \citep{Dicke1954, Houde2019}. Studying this effect requires bursts with high signal-to-noise ratios ($S/N$) and consistent dispersion measure (DM). Therefore, large numbers of bursts, which are close in time, are necessary to get a consistent picture of the effect.

In this paper, we present our statistical analysis of 849 new bursts from FRB~121102 observed with the Arecibo Telescope. We found these bursts in 10 observations during \ed{the \textit{November rain},} a burst storm in October and November 2018, \ed{where \citet{Cruces2021} already found high activity in one observation with the Effelsberg telescope}. The sample was collected using new RFI excision techniques and an improved search pipeline. Apart from the general statistical properties, we do an in-depth analysis of the wait-times, energies, and intra-burst drifts.

We describe our observational campaign, the details of our observing system, and our FRB search pipeline in Section~\ref{sec:obs}. In Section~\ref{sec:fitting} we describe the extraction of burst properties, our two-dimensional Gaussian fits, and the energy calculation. Section~\ref{sec:results} presents our analyses and results, which we discuss further in Section~\ref{sec:discussion}. In Section~\ref{sec:conclusion} we summarize and conclude.

\section{Observations and Search}
\label{sec:obs}
We observed FRB~121102 with the 305-m William E.\ Gordon radio telescope at Arecibo as part of our regular monitoring campaign at 1.4 and \SI{4.5}{\GHz} (project P3054; PI: L.\ Spitler). The 10 observations presented here were all carried out with the L-Wide receiver with a nominal frequency range of \num{1150}--\SI{1730}{\MHz}, dual linear polarizations, a gain of $\sim$\SI{10.5}{\K\per\jansky}, and a system temperature of about \SI{30}{\K}\footnote{\url{http://www.naic.edu/~astro/RXstatus/Lwide/Lwide.shtml}}. The observations were scheduled based on the telescope availability and not on prior knowledge of the source's activity. They include all L-band observations of our campaign between 2018-08-26 and 2019-02-21. We used the pulsar backend PUPPI\footnote{\url{http://www.naic.edu/puppi-observing/}} to record our filterbank data coherently dedispersed to $\DM = \SI{557.0}{\dmu}$. The dedispersion is done on 8 GPU nodes, each dedispersing a frequency band of \SI{100}{\MHz} with 64 channels\footnote{\ed{The backend frequency band extends beyond the receiver band causing the mismatch between the bandwidths.}} \citep{Duplain2008}. The bands are then merged together and stored as \textsc{psrfits} files. Occasionally, one of the GPU nodes was overloaded and wrote 0s for a short time; we refer to this as \textit{dropouts}. This affected 0.68 to \SI{3.82}{\%} of the data in each of the 10 observations presented here. The data are recorded in the resulting \SI{1.5625}{\MHz} channel width, in \SI{8}{bit}, and with a sampling time of \SI{10.24}{\us}. Albeit we record the full Stokes data, we only use the total intensity, as the linear polarization is not measurable. \ed{The reasons are that the source's high rotation measure causes intra-channel Faraday rotation smearing for the given channel widths and frequencies \citep{Michilli2018}, and additionally that the source is less polarized at L-band \citep{Plavin2022}.}

We made major changes to the pipeline that was used for previous searches \citep{Gourdji2019, Hewitt2021} and which has been described in detail in \citet{Michilli2018b}.
To save disk space and speed up the search, the data was converted to total intensity and downsampled in time by a factor of 8 to a resolution of \SI{81.92}{\micro\second} using \texttt{psrfits\_subband}\footnote{\url{https://github.com/demorest/psrfits_utils}}. In contrast to previous searches, the frequency resolution was kept at \SI{1.5625}{\mega\hertz} per channel to allow for later masking of channels. DM smearing within a channel is caused by the difference $\Delta\DM$ between the real DM and the DM used for coherent dedispersion. For our channel width $\delta\nu$ it can be approximated as
\begin{equation}
\Delta t \approx a\,2\,\delta\nu\frac{\Delta\DM}{\nu^{3}} = \SI{13}{\us} \frac{\Delta\DM}{\si{\dmu}}\frac{\nu^{-3}}{\si{\per\GHz\cubed}}\,,
\end{equation}
where $\nu$ is the frequency and \ed{$a=\SI[separate-uncertainty=false]{4.1488064239(11)}{\GHz\squared\cm\cubed\per\parsec\ms}$} the dispersion constant \citep{Kulkarni2020}.
With $\Delta\DM\sim\SI{6}{\dmu}$ (see Section~\ref{sec:fitting}) it varies between \SI{15}{\us} at the top and \SI{51}{\us} at the bottom of the band. It is therefore always smaller than the resolution used in the search.

We wrote our own program \textsc{fix\_gpu\_dropouts} to identify GPU node dropouts and replace them by random data. These dropouts manifest as $1/8$th of the band going to 0 for typically $<\SI{1}{\s}$, often multiple times in a row. In previous searches, the dropouts were treated in the same way as external radio-frequency interference (RFI). This has caused many false candidates in previous versions of the pipeline, rendering parts of the data unusable. 
The same program was used to flag channels that were affected by narrowband RFI.

\textsc{fix\_gpu\_dropouts} works in three steps. Throughout the program, only the 384 channels \ed{from 1150 to} \SI{1750}{\MHz} are considered. Other channels are set to 0, this completely includes the lowest GPU band. (i) The program first identifies timestamps where all the samples across a GPU are 0 at the same time and adds \SI{8}{\ms} (100 samples) before and \SI{24}{\ms} (300 samples) after this time to account for \textit{dropping} and \textit{ramping-up} of GPU nodes. The data around the dropouts is scaled back to the mean of 96 (the \SI{8}{bit} data allows values from 0 to 255) and standard deviation of 32 that is also given to \texttt{psrfits\_subband} in the downsampling. The dropout is replaced by random Gaussian data with those same mean and standard deviation. (ii) The RFI exclusion is based on the number of outliers, which we define as samples that are $5$ times the standard deviation above the median. To identify channels often affected by RFI, we count the outliers in each channel and block of \SI{0.3}{\s} length. We ignore times with dropouts or where many channels have outliers due to broadband RFI. For the remaining time, we calculate the median number of outliers in a channel as well as the interquantile range. We set channels to 0 that have more outliers than the median plus $0.13$ times the difference between the \SI{7}{\percent} quantile and the median (as an unbiased measure of the scatter in the data); the factor was determined empirically. (iii) Ignoring RFI channels, we search for shorter dropouts, where the data values drop in a GPU node, but do not go down to 0. For each timestamp, we take the mean in frequency across a GPU and identify short dropouts where this mean drops below its median minus 6 times its standard deviation. We again replace these short dropouts by Gaussian random data. The cleaned data is stored as a new \textsc{psrfits} file. In our 10 observations, we replace 0.68--\SI{3.82}{\%} of the data and mask 69-130 channels out of 384 in the receiver band, yielding an effective bandwidth of 492.1875--\SI{396.875}{\MHz}.

We searched the data using \textsc{presto}'s \texttt{single\_pulse\_search.py}\footnote{\url{http://www.cv.nrao.edu/~sransom/presto/}} \citep{Ransom2001} in a \DM{} interval from \SI{461}{\dmu} to \SI{661}{\dmu} in steps of \SI{1}{\dmu} and with a $S/N$ limit of 5 (compared to 6 used in earlier studies). 
We grouped together events within \SI{20}{\ms} and \SI{5}{\dmu}. 
We set liberal conditions on candidates to be visually inspected: \ed{one of the events must have a \DM{} between \SI{551}{\dmu} and \SI{581}{\dmu}, and either one of the events must have a $S/N>6$ (including single event groups) or the group must consist of at least 3 events}. These liberal conditions are enabled by the earlier rigorous dropout and RFI excision. This is unlike in previous pipeline versions, where much of the RFI exclusion was done right before candidate inspection.

We visually inspect the spectra of candidates in three windows with different time spans and resolutions to classify them as real, RFI, or ambiguous.
For later analysis, we cut out real and ambiguous bursts from the full resolution data, although for the analysis in this paper we use only bursts classified as real. Our experience with the data suggests that we are able to detect the majority of bursts above a $S/N$ of about 6.

For future applications of our pipeline, we compare our classification with the one from the machine learning classifier \textsc{fetch} \citep{Agarwal2020}. We apply \textsc{fetch} on the observations with MJD 58435, 58439 and 58450. We find that of the eleven available models, model H misses the least amount of bursts. Nonetheless, when we use model H and a probability threshold of \SI{50}{\percent}, we find that \SI{19}{\percent} of the bursts in those datasets are missed. Of the bursts that are missed, \SI{90}{\percent} have a $S/N$ below 8.0 -- the minimum $S/N$ of simulated FRBs on which \textsc{fetch} was trained \citep{Agarwal2020}. For the bursts that have $S/N < 8$, \SI{34}{\percent} have a probability less than \SI{50}{\percent} (model H). \textsc{fetch} finds \SI{96}{\percent} of the bursts that have $S/N\geq 8$ and the ones that are missed are bursts that are strongly affected by (broadband) RFI. The fact that \textsc{fetch} otherwise agreed with our classification for $S/N\geq 8$ affirms that the manual inspection is reliable. We therefore continue with our manual visual inspection for the classification and do not use the classification as provided by \textsc{fetch}. We note that it would be possible to re-train \textsc{fetch} using data from the Arecibo Telescope, which could improve the performance of the classification. That, however, is beyond the scope of this paper.

\begin{table*}
	\centering
	\caption{The 10 reported observations. The ID is used in the burst names to indicate the observation. $k$ and $r$ are the parameters of a Weibull fit to the wait-times in Section~\ref{subsec:rates}. The measured $k$s are consistent with $k=1$ for which the Weibull distribution reduces to a Poisson model.}
	\label{tab:obsis}
	\begin{tabular}{llSSSSSc}
		\hline
		\hline
		\text{Date} & \text{ID} & \text{Start MJD (topocentric)} & \text{Duration / \si{\second}} & \text{Bursts} & \text{Rate / hour} & $k$ & \text{$r$ / hour}\\
		\hline
		2018-10-18 & A & 58409.346424 & 2466 & 5 & 7	(	3	) & 5.11	$_{-	1.92	}^{+	3.11	}$&	6.2	$_{-	0.7	}^{+	0.8	}$	\\
		2018-10-22 &  & 58413.336181 & 3056 & 0 & 0		 & 	$\text{--}$	 & 	$\text{--}$	\\
		2018-10-27 &  & 58418.291134 & 5002 & 0 & 0		 & 	$\text{--}$	 & 	$\text{--}$	\\
		2018-11-10 & B & 58432.259086 & 4881 & 203 & 150	(	11	) & 0.97	$_{-	0.05	}^{+	0.06	}$&	$126.1_{-10.1	}^{+	9.3	}$	\\
		2018-11-13 & C & 58435.273426 & 2969 & 180 & 218	(	16	) & 1.05	$_{-	0.06	}^{+	0.06	}$&	193.0$_{-	14.4	}^{+	15.4	}$	\\
		2018-11-17 & D & 58439.236887 & 5058 & 227 & 162	(	11	) & 0.96	$_{-	0.05	}^{+	0.05	}$&	154.5$_{-	10.5	}^{+	10.9	}$	\\
		2018-11-20 & E & 58442.252477 & 2980 & 49 & 59	(	8	) & 0.85	$_{-	0.10	}^{+	0.11	}$&	50.5$_{-	8.5	}^{+	10.3	}$	\\
		2018-11-22 & F & 58444.24022 & 1503 & 17 & 41	(	10	) & 1.43	$_{-	0.27	}^{+	0.33	}$&	37.2$_{-	5.8	}^{+	7.5	}$	\\
		2018-11-26 & G & 58448.241505 & 2095 & 67 & 115	(	14	) & 0.89	$_{-	0.08	}^{+	0.09	}$&	110.3$_{-	13.9	}^{+	17.0	}$	\\
		2018-11-28 & H & 58450.232199 & 2815 & 101 & 129	(	13	) & 1.06	$_{-	0.08	}^{+	0.09	}$&	121.8$_{-	11.0	}^{+	11.5	}$	\\
		\hline
	\end{tabular}
\end{table*}

\begin{figure*}
	\includegraphics[width=\textwidth]{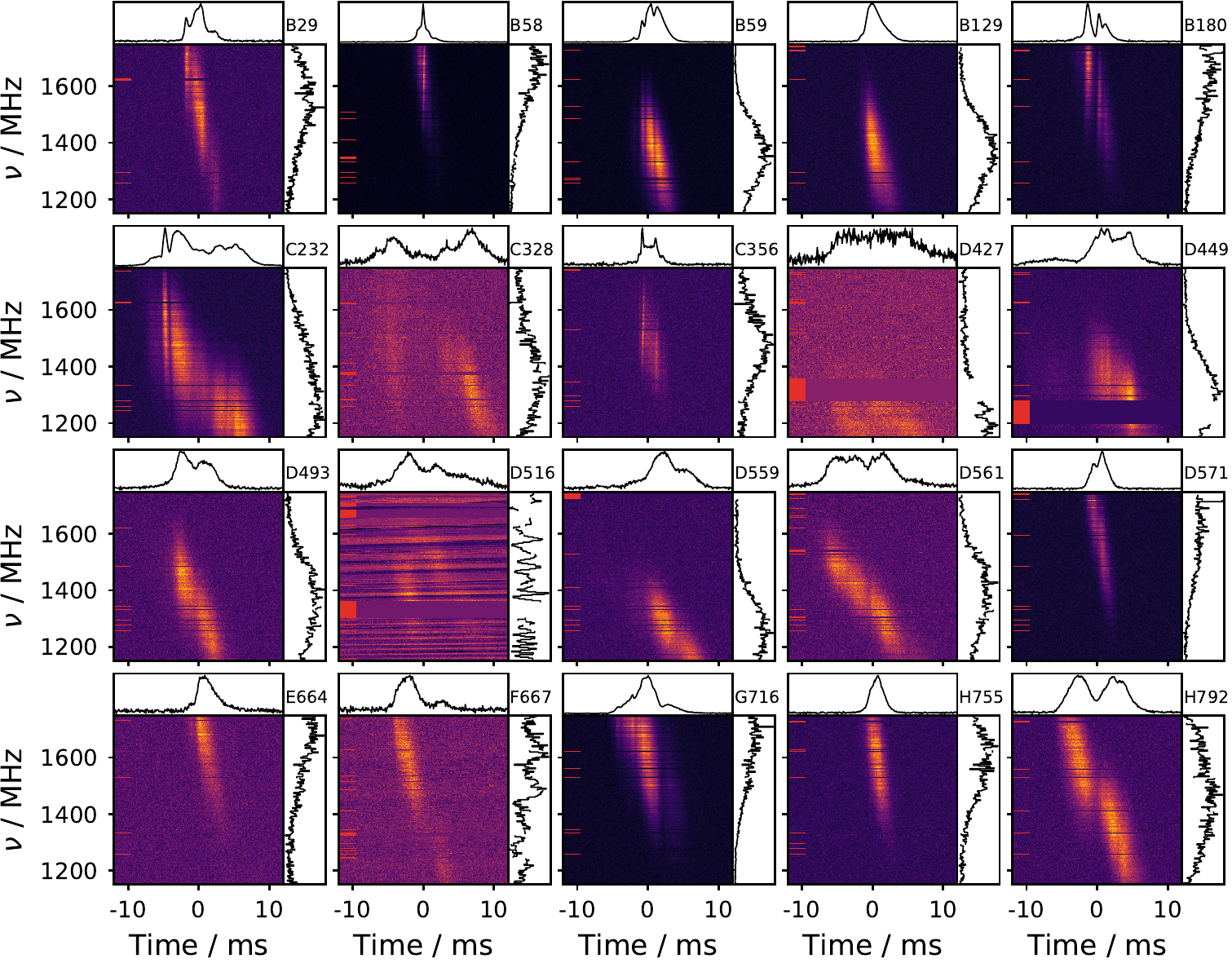}
	\caption{The 20 most energetic bursts. For each burst, the dynamic spectrum is shown with the time series (averaged over the full observing band; top panel) and the spectrum (averaged over the plotted \SI{24}{\ms} window; right panel). The burst IDs are shown in the top right corner; bursts are numbered by the order of occurrence, and letters indicate the observation. All bursts have been dedispersed to a DM of \SI{563.02}{\dmu}. Red bars on the left of the dynamic spectra mark channels that have been excluded because of RFI or GPU node dropouts. The native resolution has been reduced by a factor of 8 in time and 2 in frequency (to \SI{81.92}{\ms} and \SI{3.125}{\MHz}) for visualization purposes. Almost all bursts show temporal structure as well as the sad-trombone effect.}
	\label{fig:gallery}
\end{figure*}

\section{Burst properties}
\label{sec:fitting}

To analyse the bursts, we dedispersed all of them at a common \DM. This is to be preferred because the majority of bursts does not have sufficiently resolved structure to determine an accurate DM. Moreover, \ed{844 of the 849 bursts} are observed within 18 days and \citet{Hessels2019} have found that the DM of FRB~121102 varies by only \SI{.07}{\dmu} (standard deviation) on timescales of weeks. To determine the DM, we used \texttt{DM\_phase}\footnote{\url{https://github.com/danielemichilli/DM_phase}} \citep{Seymour2019}, a code that maximizes the temporal structure in a burst rather than the peak $S/N$. The latter is not reliable, as it can be greatly affected by the sad-trombone effect. We used 8 bursts that had temporal structure and high $S/N$, such that \texttt{DM\_phase} yields small uncertainties in DM. The mean (weighted by the squared uncertainties) of \SI{563.02(22)}{\dmu} was used to dedisperse all bursts. The 8 bursts used are B29, B58, B59, B129, B180, D493, D561, and G716, which are all shown in Fig.~\ref{fig:gallery} \ed{as part of the collection of the 20
	brightest bursts. The} DM of bursts C232 and C356 could only be obtained at a later stage and would have changed the DM slightly to \SI{562.97(18)}{\dmu}. To be consistent, we continued using $\DM=\SI{563.02}{\dmu}$. The DM values that were used are shown in the top panel of Fig.~\ref{fig:rate}. The sharpest bursts B58 and C356 are visually over-dedispersed, and we will show later that it is better to only consider the sharpest bursts to find the best DM for dedispersion.

For each burst, we masked channels by eye that were contaminated by RFI and selected a time window that fully contains the burst. The on-burst region was used for fitting, while the off-burst region was used to determine the noise to normalize channels. We located the centre of each component of a burst by eye and used it as the initial value for the following fits.

We fit two-dimensional, elliptical Gaussians to each sub-burst in the burst spectra. The exact form that we fit depending on time $t$ and radio frequency $\nu$ is 
\begin{equation}
\mathcal{G}_\mathrm{2D}(t,\nu)=A \exp\left(-\frac{(t-t_0-d_t(\nu-\nu_0))^2}{2\sigma_t^2}-\frac{(\nu-\nu_0)^2}{2\sigma_\nu^2}\right),
\label{eq:2Dgaussian}
\end{equation}
with the $6$ free parameters: amplitude $A$, time of arrival $t_0$, linear temporal drift $d_t$, $\sigma_t$: duration at $\nu_0$, central frequency $\nu_0$, and bandwidth $\sigma_\nu$. This is different from the commonly used form where a Gaussian is rotated by an angle. This parametrization has the advantage that $\sigma_t$ and $\sigma_\nu$ are independent of the \DM, while $d_t$ is closely related to it. The formerly used angles $\alpha$ \ed{\citep[see e.g.{} the definition in Appendix A of][although named $\theta$ there]{Chamma2021}} can be converted into a drift rate with a small angle approximation $d_t=-\alpha\,\si{\ms\per\MHz}$ if the time and frequency units \si{\ms} and \si{\MHz} were used to obtain $\alpha$. A second parametrization and interpretations of both are discussed in Section~\ref{subsec:tilt} and further in Section~\ref{sec:discussion}. Additional forms of the elliptical Gaussian, illustrations and explanations, as well as the full conversion equations can be found in Appendix~\ref{app:gaussians}.

For the fits we used least-square fitting with the Levenberg-Marquardt algorithm\footnote{\url{https://docs.scipy.org/doc/scipy/reference/generated/scipy.optimize.least_squares.html}}. The number of sub-bursts was decided by eye and from correlations in the free parameters if sub-bursts were too close in time-frequency space. A few sub-bursts were marked as \textit{diffuse} (e.g.{} the diffuse background component in C356) and excluded from the analysis in Section~\ref{subsec:tilt}.
234 sub-bursts from 183 bursts were fit reasonably well, e.g.\ the amplitude was larger than its uncertainty. These were generally the high $S/N$ or single component bursts. For the rest we fit the time series and the spectrum with one-dimensional Gaussians to obtain arrival times, central frequencies, durations, and bandwidths. Here the spectrum was computed from the $2\sigma_{t,\mathrm{1D}}$ before and after the arrival time, where $\sigma_{t,\mathrm{1D}}$ is the $1\sigma$ width of the Gaussian in the time series.

We computed fluence and energy with the radiometer equation \citep[see e.g.][]{Lorimer2004} using a system equivalent flux density (SEFD) that depends on frequency and zenith angle $\theta$. The radiometer equation for the flux density $\mathcal{S}$ takes the form 
\begin{equation}
\mathcal{S}=\left\langle\frac{(S/N)_\mathrm{chan}\cdot \mathrm{SEFD}(\nu, \theta)}{\sqrt{n_\mathrm{p}\cdot\delta t\cdot\delta \nu_\mathrm{chan}}}\right\rangle_\nu,
\label{eq:flux}
\end{equation}
with the signal-to-noise ratio in a time sample and channel $(S/N)_\mathrm{chan}$, the number of polarizations $n_\mathrm{p}$, the sample time $\delta t$, and the frequency width of a channel $\delta\nu_\mathrm{chan}$; the average over $\nu$ goes over the full observing band. The frequency and zenith angle-dependent SEFD was calculated from system performance measurements\footnote{\url{http://www.naic.edu/~phil/sysperf/sysperfbymon.html}}. We obtained seven different $\theta$ dependent polynomials for seven different frequency bands. Generally, the observing system is less sensitive at lower frequencies and larger zenith angles. A detailed description can be found in \citet{Hewitt2021}. To get the fluence $\mathcal{F}$ we integrated $\mathcal{S}$ over the time window that was also used for fitting; in other words, we sum over the time samples $\mathcal{F}=\sum_\mathrm{samp}\mathcal{S}\,\delta t$. The energy was obtained from the known redshift $z=0.193$ \citep{Tendulkar2017}, the resulting luminosity distance $D_L=\SI{949}{\mega\parsec}$ \citep[based on the parameters obtained by][]{Planck2016}, and the observing bandwidth $\Delta\nu_\mathrm{band}$ as
\begin{equation}
E = 4\pi D_L^2 \frac{\mathcal{F}\,\Delta\nu_\mathrm{band}}{1+z}\,, \label{eq:energy}
\end{equation}
where the factor $1+z$ accounts for the redshifted burst duration \citep{Zhang2018}.

Some studies \citep[e.g.][]{Gourdji2019} use the burst bandwidth instead of the observing bandwidth, and \citet{Aggarwal2021b} argued that this is the only correct way. Yet, both methods are  correct (and equivalent) if and only if the same bandwidth is used that is also used to calculate the flux density (Equation~\ref{eq:flux}). If one makes assumptions about the spectrum and that the observing band is small compared to the central frequency, one can use the observing frequency instead of the bandwidth \citep{Li2021}. This does not seem well motivated given that the bursts of FRB~121102 are band-limited. We therefore use the observing bandwidth as described above.

The energy obtained from Equation~\ref{eq:energy} incorporates only the energy inside the band.
The band-limited nature of bursts allows us to scale the energies -- computed via Equation~\ref{eq:energy} -- with the fitted Gaussians as suggested by \citet{Aggarwal2021b}. To minimize other effects coming from the limited observing bandwidth, we excluded sub-bursts with $\nu_0$ close to the edges or beyond, i.e.\ we require $\SI{1200}{\mega\hertz}<\nu_0<\SI{1700}{\mega\hertz}$. This and the exclusion of bursts that were not well fit leave us with 746 of the 849 bursts.
We infer the full energy of a burst from the measured energy and the fitted Gaussians. The volume under a Gaussian is given by $V=2\pi A\sigma_t\sigma_\nu$. The volume in the band can be calculated as 
\begin{align}
V_\mathrm{band} &= \sqrt{2\pi} A\sigma_t\int^{\nu_\mathrm{top}}_{\nu_\mathrm{bot}}e^{-\frac{(\nu-\nu_0)^2}{2\sigma_\nu^2}}\dd{\nu} \\
&= 2\pi A\sigma_t\sigma_\nu\frac{1}{2} \left(\erf\left(\frac{\nu_\mathrm{top}-\nu_0}{\sqrt{2}\sigma_\nu}\right) - \erf\left(\frac{\nu_\mathrm{bot}-\nu_0}{\sqrt{2}\sigma_\nu}\right)\right)\,,
\end{align}
where $\nu_\mathrm{bot}$ and $\nu_\mathrm{top}$ denote the bottom and top frequency of the band and $\erf$ is the error function. Indexing the sub-bursts of a burst with $i$ we get the total energy of a burst
\begin{equation}
E_\mathrm{tot}=E\frac{\sum_i V_i}{\sum_i V_{\mathrm{band},i}}\equiv E s\,,
\label{eq:Etot}
\end{equation}
where we introduced the scale factor $s$.
In the 746 bursts, on average \SI{92}{\percent} of $E_\mathrm{tot}$ were inside the band ($\langle s^{-1} \rangle=0.92$). The lowest fraction in the band was \SI{60}{\percent}.

This method is more accurate than directly using the measured energies, but it has a few caveats that we will mention here, along with the numbers of bursts that were affected. For 52 complex bursts that could only be fit in 1D,  we made the approximation that $d_t$ is small and used $\sigma_{t,\mathrm{1D}}$ instead of $\sigma_t$. To not mix 1D and 2D fits in one burst, we used 2D fits of weak sub-bursts in 22 bursts, even though the uncertainties were high. We confirmed by eye that these were fit reasonably well, i.e.\ the fits were in the same location as the visible sub-burst and had flat residuals. Lastly, complex bursts that have single sub-bursts whose centres are outside the band are biased slightly low because their energy was attributed to the other sub-bursts, i.e.{} their energy in the band is considered, but not scaled adequately.

\section{Results}
\label{sec:results}
\subsection{Rates and wait-times}
\label{subsec:rates}

\begin{figure*}
	\begin{center}
		\ed{\includegraphics{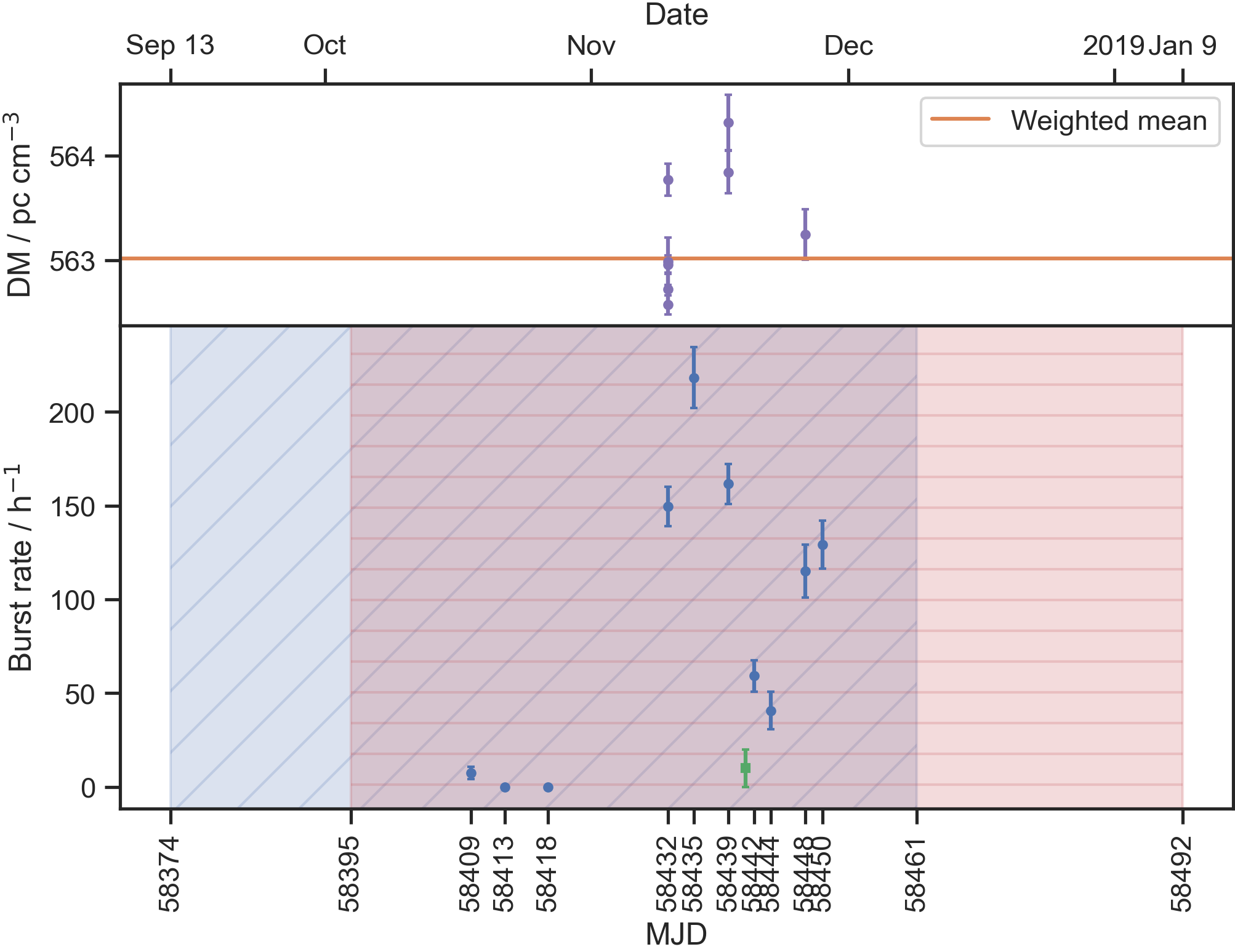}}
		\caption{Lower panel: the burst rate in each observation (see also Table~\ref{tab:obsis}). \ed{Blue, vertically hatched and red, horizontally hatched} regions mark the active cycles reported by \citet{Rajwade2020} and \citet{Cruces2021} respectively. A \ed{green} square marks the only detection by CHIME/FRB until now \citep{Josephy2019}. Upper panel: the DMs whose mean was used for dedispersion. It is shown later that the seeming DM variation is due to an emission effect.}
		\label{fig:rate}
	\end{center}
\end{figure*}
\begin{figure*}
	\begin{center}
		\includegraphics{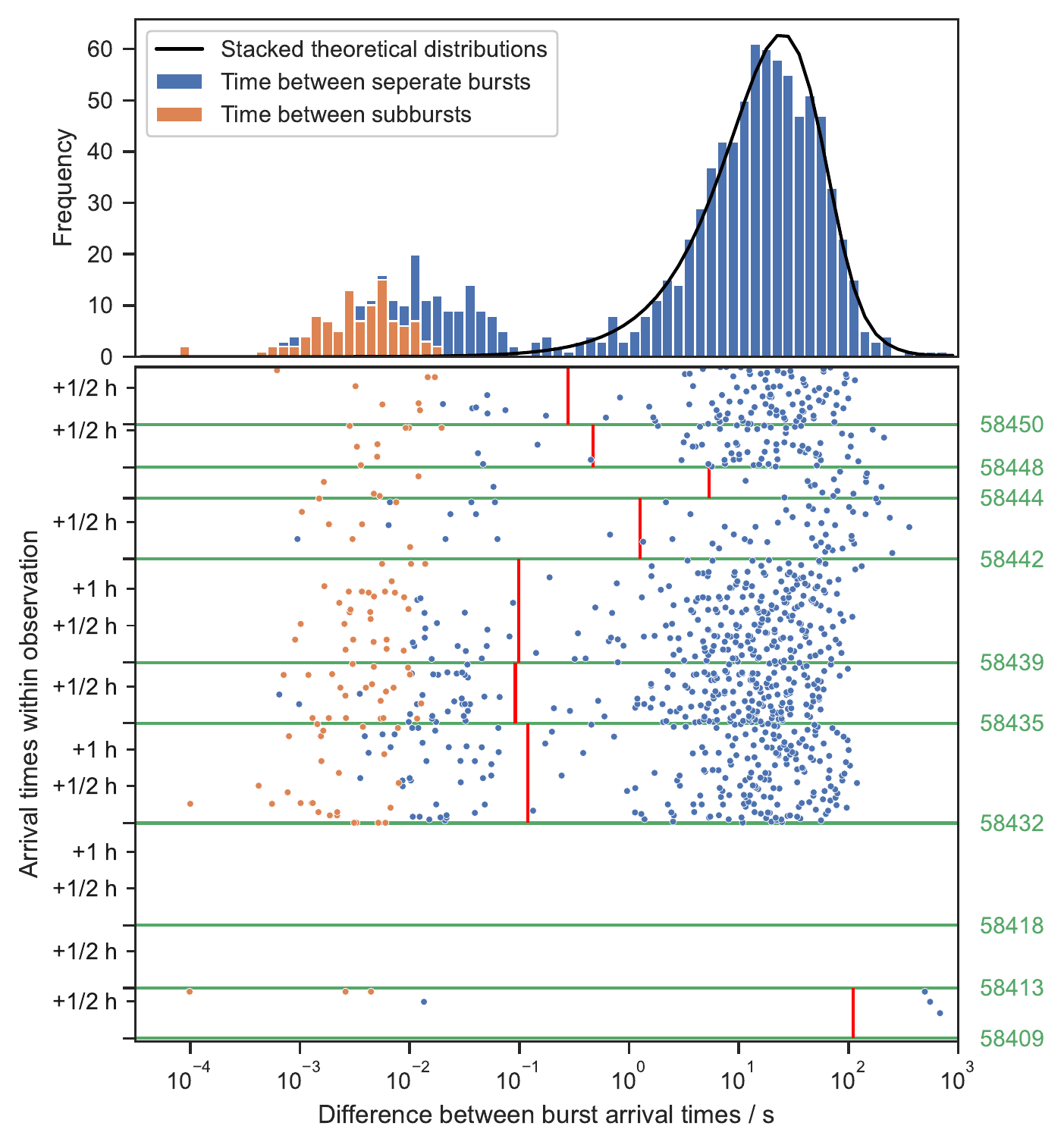}
		\caption{The wait-times between bursts and sub-bursts. Times between sub-bursts that are visually part of one burst are plotted in orange, others in blue. The lower panel has time on the y-axis but is broken at green lines, such that it only includes times when observations took place. The observation days are indicated at the start of the observations by their MJD (in green). Red lines mark the time below which only one burst is expected from Poisson statistics, given the number of bursts and observation lengths (see text for details). The upper panel shows a stacked histogram of wait-times in all observations. The black line shows the expected distribution for the right peak, assuming Poisson statistics with a different rate in each observation.}
		\label{fig:wait_times}
	\end{center}
\end{figure*}

The burst rate of each observation is listed in Table~\ref{tab:obsis} and plotted in Fig.~\ref{fig:rate}. The plotted uncertainties of the rate $r$, in an observation with burst number $N_\mathrm b$ and length $t_\mathrm{obs}$ are calculated as $\sigma(r)=\sqrt{N_\mathrm b}/t_\mathrm{obs}$, i.e.\ assuming it is a Poisson process \citep[justified by the findings of][]{Cruces2021}. Here and henceforth we show \SI{68}{\percent} uncertainties unless otherwise stated.
The observed rate of detected bursts varies significantly between observations even though the observations are in the same active window, calculated from the periods reported by \citet{Rajwade2020} as well as \citet{Cruces2021}.
It ranges from 0 bursts during the \SI{1.4}{\hour} observation on MJD 58418 to \num{180} bursts in the \SI{\sim50}{\minute} observation on MJD 58435 yielding a rate of 218 bursts per hour. This demonstrates that on timescales larger than the observations of 1--\SI{2}{hours} the times of arrival can no longer be described by a \textit{stationary} Poisson point process, but possibly still by a Poisson point process with a varying rate \ed{\citep[see also][]{Nimmo2022}}. Notably, the high-rate observations are all in the middle of the activity window reported by \citet{Cruces2021} \ed{from Effelsberg monitoring}, while the three observations with a low rate are more towards the beginning. However, the sparse sampling does not allow a definite interpretation.

We obtain wait-times between consecutive bursts -- including sub-bursts -- by taking the differences between the barycentre corrected arrival times. The joint histogram of all observations is shown in the upper panel of Fig.~\ref{fig:wait_times}. One can see two peaks and to analyse them separately we divide the wait-times into two groups, one including wait-times $\delta<\SI{0.1}{\second}$ and one with $\delta>\SI{0.1}{\second}$, where the value \SI{0.1}{\second} was chosen by eye and will be further justified below. With this division, the left peak has a median of \SI{9.7}{\ms} and the right peak a median of \SI{17.5}{\s}. Excluding the wait-times between sub-bursts, the median of the left peak becomes \SI{22}{\ms}. When including only sub-bursts, the median is \SI{4.3}{\ms}.

The timescale of the left peak is stable over observations that have very different rates, as can be seen in the lower panel of Fig.~\ref{fig:wait_times}, where the wait-times are plotted against the arrival time within their observation. This means that the left peak must reflect a characteristic timescale of the emission process.

Furthermore, only half of the left peak consists of sub-bursts that visually belong to the same burst, while the other half consists of bursts between which the signal goes down to the noise or that were even picked up as separate bursts by the pipeline. This confirms that there is a higher chance to get bursts in close proximity but up to \SI{\sim0.1}{\second} apart. This suggests that these burst packs are part of the same physical ``event'', happening in (the vicinity of) the potential neutron star or other FRB central engine.

In contrast, the peak at higher wait-times in Fig.~\ref{fig:wait_times} reflects the inverse rate, i.e.\ it shifts to longer wait-times in observations with lower rates as one expects if burst arrival times follow a Poisson point process within observations. In a Poisson point process, bursts occur independently of one another and are solely described by their rate $r$. The wait-times $\delta$ follow an exponential probability density function $\mathcal{P}(\delta|r)=r\,\mathrm e^{-r\delta}$, with mean $\langle\delta\rangle=r^{-1}$. Since the rate varies between observations, the right peak in Fig.~\ref{fig:wait_times} can be described by the joint distribution as $\sum_i n_i \mathcal{P}(\delta|r_i)$, where $i$ goes over all observations, $n_i$ is the number of arrival times $\delta > \SI{0.1}{\s}$, and $r_i=n_i/t_{\mathrm{obs},i}$. It is plotted as a black line in the upper panel of Fig.~\ref{fig:wait_times} for the same bins that were used for the histogram. We can also calculate the separation $\delta$, below which we expect to find only one burst pair in an observation with $N_\mathrm b$ bursts. It follows from the cumulative distribution function as
\begin{align}
\mathrm{CDF}(\delta|r)=1-\mathrm e^{-r\delta} & =1/N_\mathrm b \qquad\text{and therefore}\\ 
\delta &=-1/r\log(1-1/N_\mathrm b)\,.
\end{align}
It is shown as red lines in Fig.~\ref{fig:wait_times} for the number of bursts in each observation (including ones with $\delta < \SI{0.1}{\s}$). This provides a quantitative separator between the peaks of short and long wait-times and justifies the choice of \SI{0.1}{\s} as the approximate peak separation.

To confirm that the arrival times are distributed according to Poisson statistics, we used the Bayesian analysis with a Weibull distribution developed by \citet{Oppermann2018}. In the past, the Weibull distribution was used to explain the highly non-Poisson nature of the burst arrival times. The source was seen to be highly variable on timescales of weeks to months, which is now explained by the \num{\sim160}-day periodicity. On shorter timescales, \citet{Cruces2021} have shown that arrival times within an observation are consistent with Poisson statistics if short burst separations are excluded. Here we repeat their analysis with a larger number of bursts, providing better constrains. The Weibull distribution is a generalization of the Poissonian model with a second parameter $k$ but includes the latter as a special case. Here, we use it to test if the resulting parameters are consistent with this special case of Poisson statistics. For wait-times $\delta$ the Weibull distribution can be written as $\mathcal{W}(\delta|k, r) = k\delta^{-1} [\delta r \Gamma (1 + 1/k)]^k e^{-[\delta r \Gamma(1+1/k)]^k}
$, where $\Gamma(x)$ is the gamma function, r is the rate, and $k$ is the parameter that is $>1$ for periodic arrival times, $<1$ for clustered bursts, and $1$ for bursts following Poisson statistics. We used the arrival time of burst packs by taking the mean arrival time of bursts that were separated by $<\SI{0.1}{\s}$.  We inferred the parameters of the Weibull distribution separately for each observation and obtained the posterior distributions with the formalism described in \citet{Oppermann2018}. The resulting $k$ and $r$ parameters are listed in Table~\ref{tab:obsis} and are all consistent with the Poisson model. As an example, we show the posterior distribution of the highest rate observation (MJD 58435) in Fig.~\ref{fig:weibull}. For comparison, we also show the posterior distributions for the arrival times of all bursts (i.e.\ without averaging over arrival times with $\delta < \SI{0.1}{\s}$).

\subsection{General properties}
\label{subsec:properties}

\begin{figure*}
	\begin{subfigure}{.33\textwidth}
		\includegraphics[width=\columnwidth]{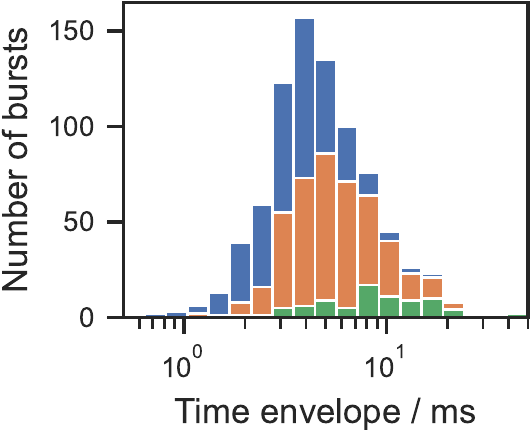}
	\end{subfigure}
	\begin{subfigure}{.33\textwidth}
		\includegraphics[width=\columnwidth]{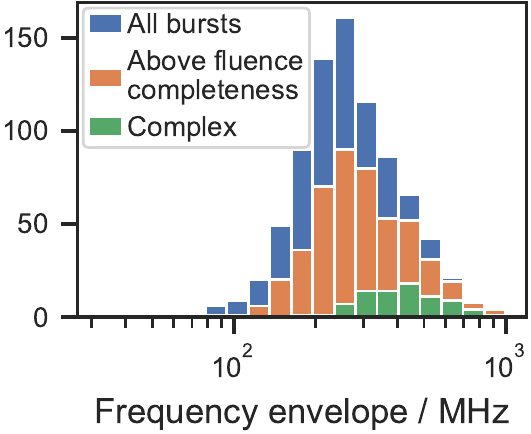}
	\end{subfigure}
	\begin{subfigure}{.33\textwidth}
		\includegraphics[width=\columnwidth]{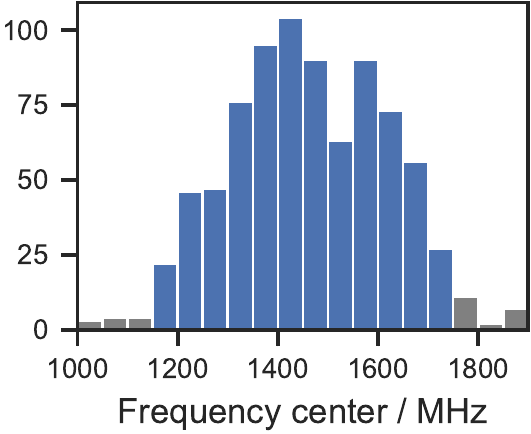}
	\end{subfigure}
	\caption{The distributions of burst FWHM durations (left) and bandwidth (middle) as well as the central frequencies (right) of the 820 bursts that were well fit. For complex bursts we took the time difference between the later half maximum of the latest sub-burst and the early half maximum of the first sub-burst and analogously for the bandwidths. The central frequencies show no clearly preferred frequency range.}
	\label{fig:envelopes}
\end{figure*}

A small sample of the detected bursts can be seen in Fig.~\ref{fig:gallery}. Like previously discovered bursts, they exhibit a variety of morphologies. Most of the bright bursts show spectro-temporal structures including multiple sub-bursts as well as the sad-trombone effect. Like in previous observations, no clear upward drifting bursts were found. All bursts are band-limited, although the sub-bursts of some drift over the full band. While most burst durations are of the order of \SI{\sim5}{\ms}, the sharpest sub-bursts can be shorter than \SI{.2}{\ms}. We show the fit parameters and additional info of the first 10 sub-bursts in Table~\ref{tab:bursts}, the full list is provided as supplementary material.

We first report the general properties of the full bursts (rather than of individual sub-bursts). This includes the distributions of duration, bandwidth, and central frequencies. We report the burst durations in terms of the full width at half maximum (FWHM) for simple bursts that were fit with a single component. For complex bursts with several sub-bursts, we compute for each sub-burst the earlier half maximum $t_0-\sigma_{t,\mathrm{1D}}\sqrt{2\ln2}$ and the later half maximum $t_0+\sigma_{t,\mathrm{1D}}\sqrt{2\ln2}$ and report the difference between the first and the last half maximum in the burst. To be consistent between bursts fit in 1D and 2D, we calculate $\sigma_{t,\mathrm{1D}}$ from the 2D fit parameters as $\sigma_{t,\mathrm{1D}}^2=\sigma_t^2+(d_t\sigma_\nu)^2$.
Similarly, we report the burst bandwidth as the FWHM, or for complex bursts the difference between the highest
frequency half maximum and the lowest frequency half maximum (the parameters in 1D and 2D are equivalent here). 
The resulting distributions are plotted in Fig.~\ref{fig:envelopes} for the full sample and with bursts above the fluence completeness threshold that we will calculate in Section~\ref{subsec:energy}. The resulting median duration of all bursts is \SI{4.41}{\milli\second} and for the bandwidth it is \SI{217}{\mega\hertz}. The more physically meaningful medians for the bursts above the completeness threshold are \SI{5.15}{\milli\second} and \SI{240}{\mega\hertz} and the first and last deciles are $\mathrm{FWHM}_{t,10\%}=\SI{2.74}{\milli\second}$, $\mathrm{FWHM}_{t,90\%}=\SI{12.54}{\milli\second}$, $\mathrm{FWHM}_{\nu,10\%}=\SI{150}{\mega\hertz}$, and $\mathrm{FWHM}_{\nu,90\%}=\SI{439}{\mega\hertz}$.

The central frequencies of bursts shown in Fig.~\ref{fig:envelopes} are subject to several biases. Only the 818 bursts whose fits gave reasonable results are shown, and for complex bursts we took the mean of the obtained $\nu_0$'s. The number of burst centres falls off at the edges because of two biases. First, bursts that are not fully in the band are less likely to be detected, and second, these bursts are more likely to have larger errors and therefore be excluded from the analysis. Additional biases come from the frequency dependent receiver sensitivity, which is overall lower at lower frequencies \citep[see fig.~7 in][]{Hewitt2021}, from RFI, and from GPU node dropouts. Without mitigating for all these effects in detail, we can say that the distribution and the median within the band of \SI{1456}{\mega\hertz} (the central frequency is \SI{1440}{\mega\hertz}) show no clearly preferred frequency band. \ed{It could be that the prefered frequency is \SI{\sim1450}{\mega\hertz}, and that the burst rate is so high because the preferred frequency is so close to the central frequency, but verifying this would require a detailed analysis of the listed biases, which is beyond the scope of this work.}

\begin{figure*}
	\begin{center}
		\includegraphics[width=1.\textwidth]{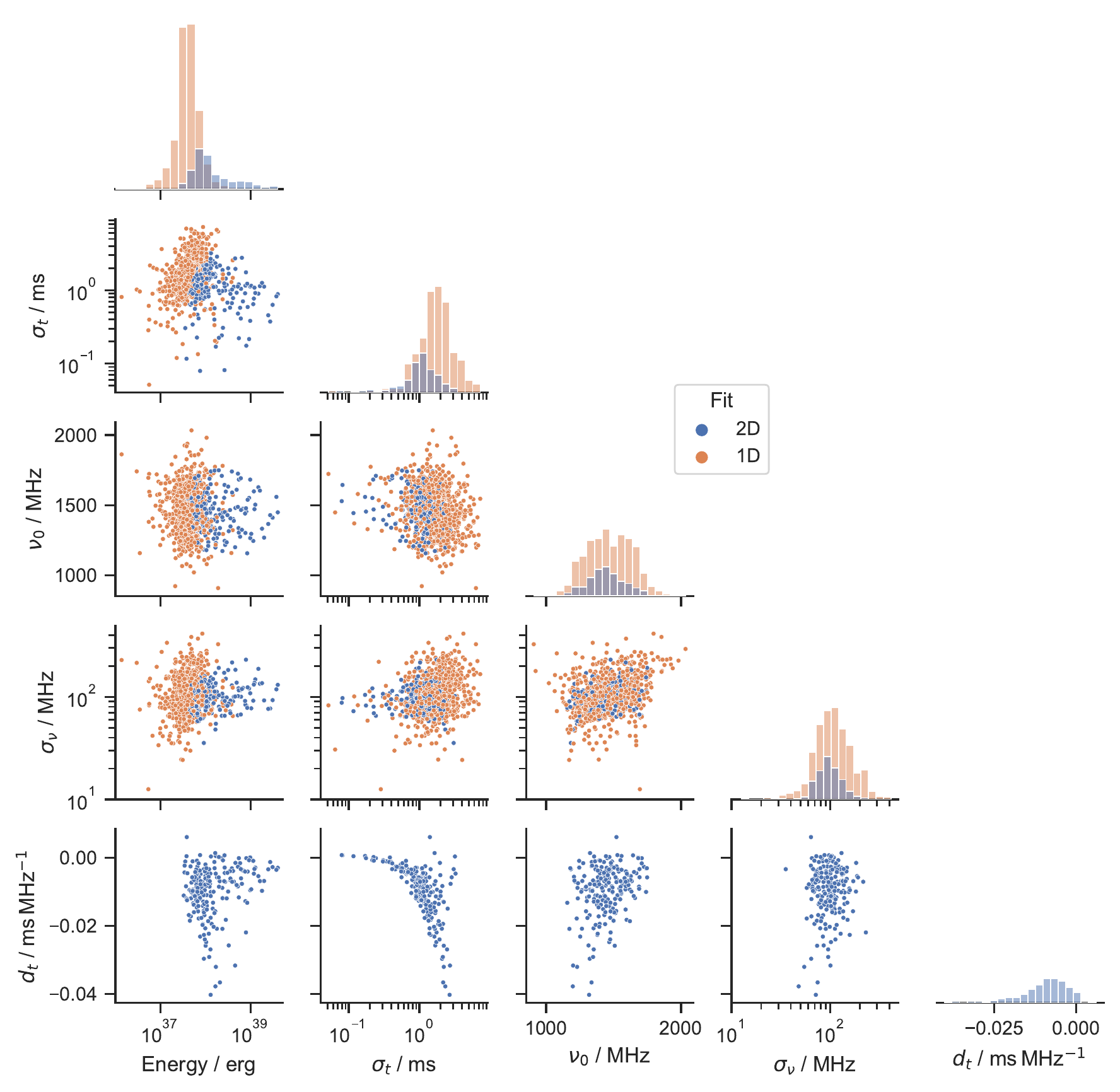}
		\caption{Corner plot of the relations between different fitted parameters and the energy. Note that $d_t$ was not fitted in the 1D model. The strongest correlation is in the $d_t$--$\sigma_t$ plot.}
		\label{fig:corner_plot}
	\end{center}
\end{figure*}

In addition to the burst envelopes, we also looked at the properties of sub-bursts. We explored the full space of quantities obtained from sub-bursts to look for unexpected correlations. Fig.~\ref{fig:corner_plot} shows a corner plot of the relevant fitted parameters and the isotropic-equivalent energy. For sub-bursts, the energy that was measured for the full bursts had to be calculated from the sub-burst parameters. We did this by distributing the measured energy among the sub-bursts through weighting by the volume under the fitted Gaussians. The tilt $d_t$ is only plotted for bursts or sub-bursts that were fit in 2D because it is not obtainable from the 1D fits. The $\sigma_t$ parameter for 1D and 2D fits are plotted in the same panels but are mathematically different because $\sigma_t$ in Equation~\ref{eq:2Dgaussian} is the width at the central frequency while $\sigma_{t,\mathrm{1D}}$ is the width in the time series. The two are related by the tilt as $\sigma_{t,\mathrm{1D}}^2=\sigma_t^2+(d_t\sigma_\nu)^2$.

The most striking relationship can be seen between $d_t$ and $\sigma_t$. We want to stress again that in the parametrization that we use in Equation~\ref{eq:2Dgaussian} $d_t$ and $\sigma_t$ are mathematically independent. We will further investigate this relationship in Section~\ref{subsec:tilt}.

Apart from this, we find weaker correlations in $d_t$--$\nu_0$ and $\sigma_\nu$--$\nu_0$.
We looked for temporal trends or systematic changes in the distributions of $\nu_0$, $\sigma_t$, and energy with time but did not find any. Bursts also showed no significant correlation between their energy and wait-times, before or after a burst.
Bursts of the same pack showed a weak tendency that later arriving bursts are at lower frequency. 63 out of 103 bursts that arrived within \SI{0.1}{\s} after another burst had a lower frequency than their predecessor, which is $2.26\sigma$ from the mean.

\ed{A positive correlation between wait-times and energy could indicate a build-up of energy that is released with the burst. The observed lack of a correlation, however, can have several reasons. Beaming effects could wash out the correlation. The energy budget of FRBs is small compared to high frequency bursts from magnetars and a built-up might not be needed. The missing wait-time–energy correlation therefore provides no clear insight into the emission mechanism.}

\subsection{The time-frequency drift}
\label{subsec:tilt}

\begin{figure}
	\includegraphics[width=\columnwidth]{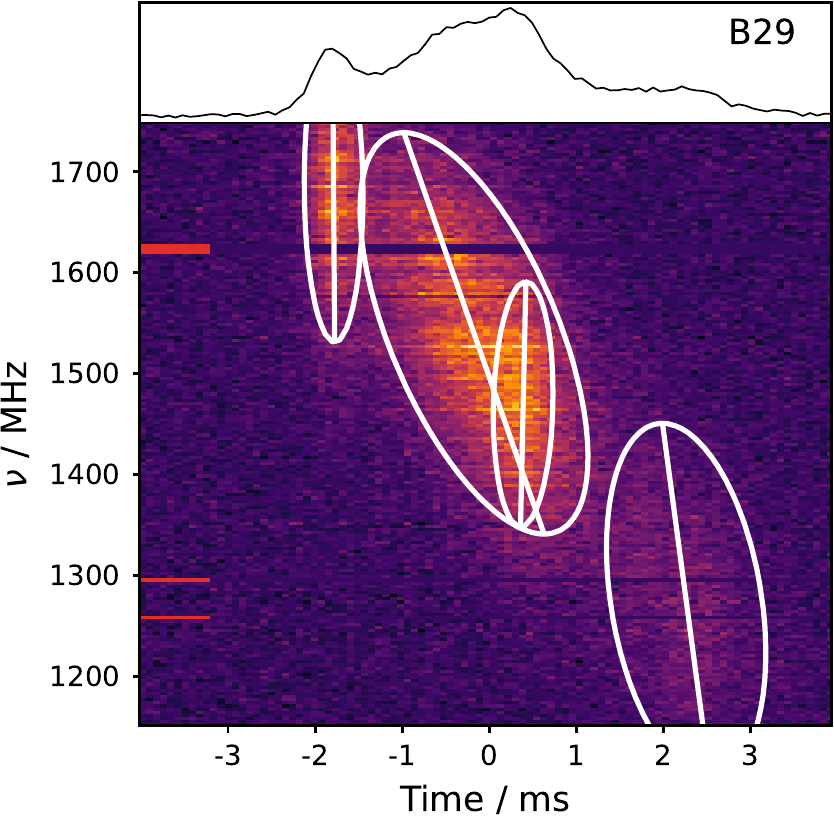}
	\caption{Example of the temporal drift $d_t$ and the $d_t$--$\sigma_t$ relationship. White ellipses show the \SI{32}{\percent} contour for each sub-burst, individually computed from the Gaussian fits. A white line shows the fitted $d_t$, i.e.\ the line along which $t_0$ drifts with frequency. It is visible that the two shorter sub-bursts drift much less than the broader components.}
	\label{fig:tilt_illustration}
\end{figure}

\begin{figure}
	\includegraphics[width=\columnwidth]{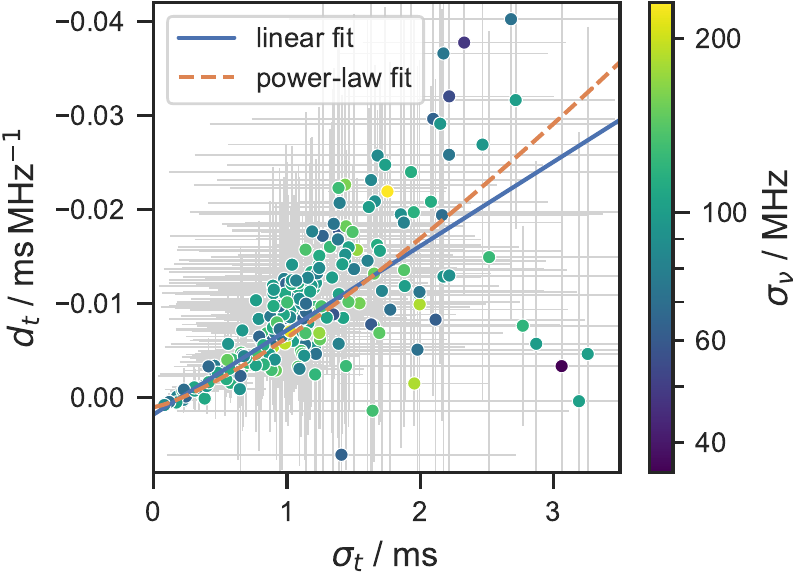}
	\caption{The intra-burst drift $d_t$ plotted against $\sigma_t$. Similar to the second panel in the bottom row of Fig.~\ref{fig:corner_plot}, but with a linear $\sigma_t$-axis and the colour representing $\sigma_\nu$ on a logarithmic scale. Uncertainties are shown in grey. The fitted linear- and power-law models are \ed{similar by eye, but not consistent with each other}.}
	\label{fig:tilt}
\end{figure}

\begin{figure}
	\includegraphics[width=\columnwidth]{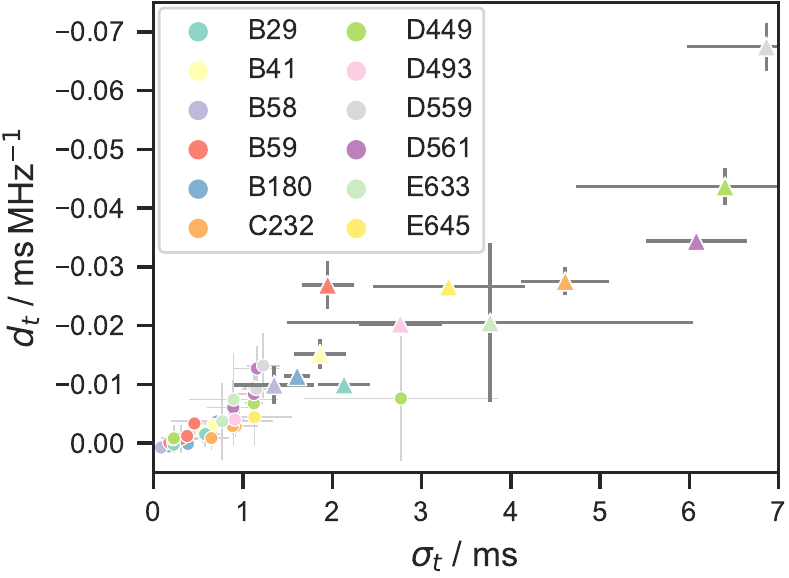}
	\caption{Comparison of the temporal drift from the sad-trombone effect (triangles) to the intra-burst drift (circles) for the 12 bursts with three or more sub-bursts. Full bursts have larger drifts, but are in the same regime as broader sub-bursts in Fig.~\ref{fig:tilt}.}
	\label{fig:tilt_full}
\end{figure} 
\begin{figure}
	\includegraphics[width=\columnwidth]{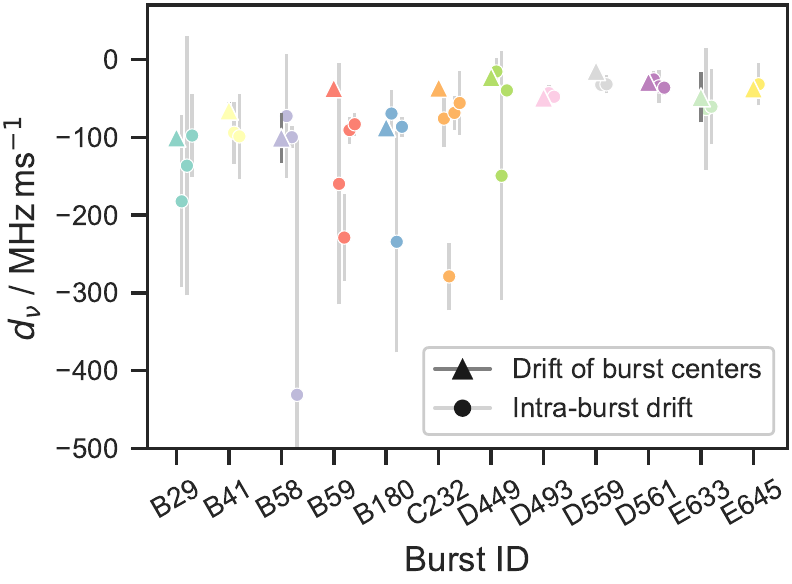}
	\caption{An alternative interpretation of the sub-burst drifts, in terms of a drift of the frequencies instead of the time centre (note the inverse units). The frequency drift of sub-burst centres due to the sad-trombone effect are mostly compatible with the intra-burst drift for the same sub-bursts. The colours are the same as in Fig.~\ref{fig:tilt_full}.}
	\label{fig:drift_full}
\end{figure}

Among the diverse spectro-temporal effects observed in FRBs, two can be well quantified: the drift between sub-bursts (i.e.{} the ``sad-trombone effect") and the intra-burst drift. To avoid confusion, we will refer to the first as sad-trombone drift. The sad-trombone effect is the effect that within bursts with multiple sub-bursts, the central frequencies drift to lower frequencies with time. This effect can be seen in most of the bright bursts in Fig.~\ref{fig:gallery}, good examples are bursts B29, B180, and D516. It is commonly quantified in the units \si{\MHz\per\ms} \citep[e.g.\ in][]{Hessels2019}. The intra-burst drift, on the other hand, is the drift of the emission within a sub-burst, it is possibly -- but not necessarily -- related to the sad-trombone drift. In the form that we fit to the bursts (Equation~\ref{eq:2Dgaussian}) lower frequencies arrive later and hence it drifts in time with frequency with (inverse) units of \si{\ms\per\MHz}. This effect is illustrated in Fig.~\ref{fig:tilt_illustration}. In the context of our Gaussian models, the sad-trombone drift can be described as the central frequencies $\nu_0$ decreasing for later arriving sub-bursts. The intra-burst drift is quantified by the fit parameter $d_t$.

Instead of letting a sub-burst drift in time (i.e.\ $t_0$ drifts) as in Equation~\ref{eq:2Dgaussian} one can equivalently let the frequency $\nu_0$ drift. Both alternatives can yield a perfectly (elliptical) Gaussian burst, such that it cannot be decided whether a sub-burst is drifting in time or in frequency. Likewise, the way we interpret the sad-trombone effect as a frequency drift is not unambiguous, in many bursts it could also be interpreted as a delay of lower frequencies. The likely reason it is perceived as the first is due to the visible substructure. Specifically, if the substructure in bright bursts were not vertical but horizontal we would interpret it as a drift in time, where bursts at lower frequencies arrive later (here ignoring the potential difficulty in DM determination). Since no clear substructure is visible in sub-bursts, we will discuss both interpretations of the intra-burst drift as well as the sad-trombone drift in this section. When needed, we distinguish them as ``temporal drift'' (in \si{\ms\per\MHz}) and ``frequency drift'' (in \si{\MHz\per\ms}). The frequency drift will show its worth as the interpretation that is closer to the sad-trombone drift, but the temporal drift can often be measured more precisely and allows us to measure and correct the DM.

In Fig.~\ref{fig:tilt} we show again $d_t$ plotted against $\sigma_t$ where we already saw a correlation in the previous section. The low scatter in the points compared to the error bars indicates that the 2D Gaussian fit to the bursts likely overestimates the uncertainties. A trend is visible between $d_t$ and $\sigma_t$ with sharper bursts having very small $d_t$, while it gets larger with longer burst durations. Above the points that follow the trend, a sharp edge is visible, which means that there is a limit on how much a sub-burst can drift for a given $\sigma_t$. Below the trend, the edge is less sharp and a few outliers with positive $d_t$ exist that are all weak single component bursts whose measurement is likely affected by unresolved sub-bursts.

A change in the intra-burst drift could be wrongly ascribed to a varying DM, but the close relationship with $\sigma_t$ strongly suggests that it must be an effect intrinsic to the emission mechanism, as a variation in the DM would add to $d_t$ independent of $\sigma_t$. Furthermore, a relation between $d_t$ and the duration measured in the time series ($\sigma_{t,\mathrm{1D}}$) would be expected naturally, but we want to stress again that it is not the origin of the correlation here because $\sigma_t$ in Equation~\ref{eq:2Dgaussian} is instead the width of a sub-burst at the central frequency $\nu_0$.

From the illustration in Fig.~\ref{fig:tilt_illustration} we can already partly understand the $d_t$--$\sigma_t$ relationship. If we imagine that the second sub-burst consisted of several unresolved smaller sub-bursts with low intra-burst drift (like the first and third), the sad-trombone drift of their central frequencies causes the larger $d_t$ in the wider (larger $\sigma_t$) observed sub-burst. 
We will see after the following quantitative analysis that this simple picture cannot completely explain the relationship.

To quantify the $d_t$--$\sigma_t$ trend \ed{in absence of a theoretical model}, we fit a power-law \ed{and a straight line} to the data. \ed{The power-law} is of the form \ed{$d_t(\sigma_t)=-(b\,\sigma_t)^k\si{\ms\per\MHz}+c$}. We use the Levenberg-Marquardt algorithm \ed{to minimize the sum of the squared residuals and weight by the squared uncertainty of $d_t$. We find 
	$k=\num{1.28(8)},\,b=\SI{.022(5)}{\per\ms},\,\text{and}\,c=\SI{0.0010(1)}{\ms\per\MHz}$. 
	When fitting a linear $d_t(\sigma_t)=b\,\sigma_t+c$, we get $b=\SI[separate-uncertainty=false]{-0.00862(37)}{\per\mega\hertz}$ and
	$c=\SI[separate-uncertainty=false]{0.00171(30)}{\milli\second\per\mega\hertz}$. The power-law fit does not agree with a linear law, but visually it does not describe the data much better, because of the large scatter in the data. For the following analysis, we will therefore use the simpler linear model.}

We want to compare the intra-burst drift to the sad-trombone drift. To do this, we calculate the sad-trombone drift for the 12 bursts that have three or more sub-bursts by fitting a line to the sub-burst centres. One can choose to interpret the resulting slope as a temporal drift in frequency in units of \si{\ms\per\MHz} or inversely as frequency centres drifting with time in \si{\MHz\per\ms}. We first compare the temporal drifts to see if there is a difference in magnitudes between the sad-trombone drift and the intra-burst drift. The sad-trombone drifts of the 12 bursts are shown in Fig.~\ref{fig:tilt_full} along with the $d_t$ of the sub-bursts of the same bursts. The FWHM calculated in Section~\ref{subsec:properties} has been divided by $2\sqrt{2\ln{2}}\approx2.355$ to get a quantity that is comparable to $\sigma_t$. We can see that the sad-trombone drifts are generally higher than the intra-burst drifts, which is not surprising as the sub-bursts already appear straighter in their spectra in Fig.~\ref{fig:gallery}. Moreover, the bursts seem to continue the same trend without a gap between the drifts of sub-bursts and full bursts. Rather, the sad-trombone drifts fall in the same regime as broader sub-bursts in Fig.~\ref{fig:tilt}. This is also not unexpected, as we know that weaker bursts sometimes consist of several unresolved sub-bursts.

A more suitable comparison can be done by comparing the drift of frequency centres to the drift of the emission frequency in individual sub-bursts. To do this, we first need to convert the $d_t$ to a comparable quantity. For that, the drift of the emission frequency can be obtained by converting to a different parametrization of the 2D Gaussians, where -- in contrast to Equation~\ref{eq:2Dgaussian} -- the frequencies drift linearly with slope $d_\nu$ in units of \si{\MHz\per\ms},
\begin{equation}
\mathcal{G}_{\mathrm{2D},\nu}(t,\nu)=A \exp\left(-\frac{(t-t_0)^2}{2w_t^2}-\frac{(\nu-\nu_0-d_\nu(t-t_0))^2}{2w_\nu^2}\right)\,.
\label{eq:2DGaussian_nu}
\end{equation}
Here $w_t$ reflects the full duration in the time series (i.e. is equivalent to $\sigma_{t,\mathrm{1D}}$), and $w_\nu$ is the bandwidth at $t_0$. All parameters can be calculated from the fitted parameters as done in Appendix~\ref{app:gaussians}; for $d_\nu$ one gets
\begin{equation}
d_\nu=\frac{d_t}{d_t^2+\frac{\sigma_t^2}{\sigma_\nu^2}}\,. \label{eq:drift_nu}
\end{equation}
For the sharpest bursts, for which $|d_t|\ll\sigma_t/\sigma_\nu$ still holds, Equation~\ref{eq:drift_nu} can be approximated as $d_\nu\approx d_t\sigma_\nu^2/\sigma_t^2$, and uncertainties in $d_t$ coming from the DM that are greater than $d_t$ itself can cause variations in $d_\nu$ of the order of several \SI{100}{\MHz\per\ms}. This causes large uncertainties in $d_\nu$ for the \num{\sim5} sharpest bursts and requires us to correct for the small over-dedispersion $\delta\DM$ (visible in bursts B58 and C356). We do this by subtracting each $d_t$ by a small $d_t(\nu_0)$, which is derived below in Equation~\ref{eq:dnu}. In Fig.~\ref{fig:drift_full}, we show the drifts of frequency centres and the drifts of the emission frequency within sub-bursts $d_\nu$ for the same 12 bursts as in Fig.~\ref{fig:tilt_full}. The strong susceptibility to small offsets in DM of some bursts is reflected in their large error bars. For many bursts, the sad-trombone drifts and the intra-burst drifts are very similar and within the uncertainties. Exceptions from this are bursts B59 and C232. In burst C232 in Fig.~\ref{fig:gallery} it is visible that the later sub-bursts stop drifting, which might cause the lower sad-trombone drift.

Even though most bursts agree with the sad-trombone drift, we have to note that the majority of error bars on the $d_\nu$ are relatively large. Moreover, the linear $d_t$--$\sigma_t$ relation does not agree with equal sad-trombone and intra-burst drifts in the following simple picture. Let us hypothesize a toy burst that has frequency drift $d_\nu$ and a Gaussian frequency envelope with width $w_\nu$ at any point in time. If it is then modulated in time into sub-bursts using Gaussians with varying widths, we can calculate the resulting $d_t$ from Equations~\ref{eq:Gtparams} and \ref{eq:Gnuparams}. The resulting equality $d_t=\sigma_t^2d_\nu/w_\nu^2$ is a square-law and in conflict with the observation of a linear relation.

We mentioned in Section~\ref{sec:fitting} that $d_t$ is closely related to the DM that is used for dedispersion. The reason is that if we over-dedisperse the bursts it will add a positive intra-burst drift to all of them, while under-dedispersion causes the opposite. Since we dedispersed all bursts at the same DM, this does not affect the slope $b$ but only $c$. In fact, the small positive intra-burst drift $d_t$ of the sharpest bursts in Fig.~\ref{fig:tilt} suggests that we used a slightly too high DM. Under the assumption that the linear law extends to very sharp bursts ($\sigma_t\rightarrow 0$) and for these bursts $d_t$ becomes 0, we can estimate the real $\DM$ from $c$ as follows.

If the real DM is $\DM_\mathrm{real}=\DM_\mathrm{applied}+\delta\DM$, i.e.\ it is higher than the applied dispersion by a small $\delta\DM$, it will cause a small shift $\Delta t$ with respect to a central frequency $\nu$ (in our case $\delta\DM$ will be negative because we over-dedispersed). The apparent intra-burst drift caused by the wrong DM can then be approximated with the tangent at $\nu$ by
\begin{equation}
d_t(\nu) = \frac{\Delta t}{\Delta \nu} \approx \dv{\nu}a\times\delta\DM\left(\frac{1}{\nu^2}-\frac{1}{\nu_\mathrm{ref}^2}\right) = -2a\times\delta\DM\frac{1}{\nu^3}\,.
\label{eq:dnu}
\end{equation}
A more rigorous derivation is given in Appendix~\ref{subsec:taylorDM}.
Solving for $\delta\DM$ we obtain
\begin{equation}
\delta\DM=-\frac{1}{2}\frac{\nu^3}{a}d_t \,.  \label{eq:ddm}
\end{equation}
And finally putting in $d_t=c$, and the frequency at the centre of the band $\nu=\SI{1440}{\mega\hertz}$ we get $\delta\DM=\SI{-0.61(11)}{\dmu}$ and therefore
\begin{equation}
\DM_\mathrm{real}=\SI{563.02}{\dmu}+\delta\DM=\SI{562.41(11)}{\dmu}\,.
\end{equation}

\begin{figure}
	\includegraphics[width=\columnwidth]{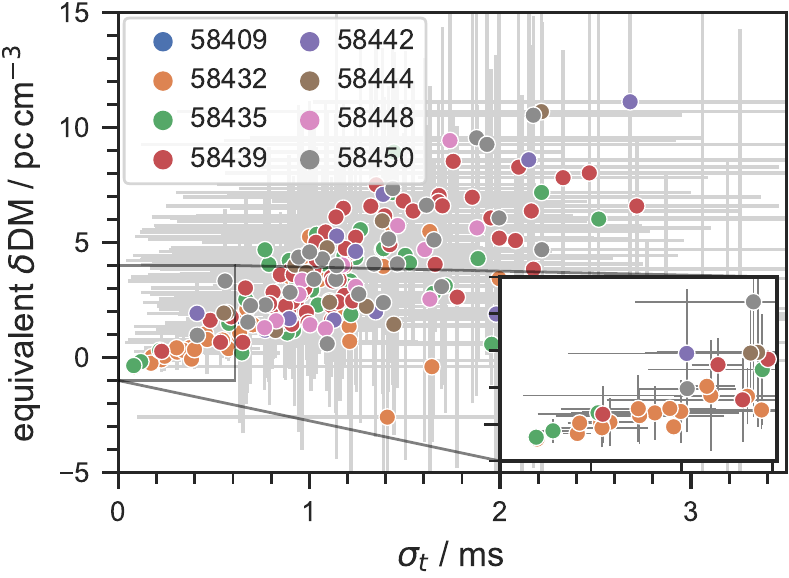}
	\caption{The apparent DM difference from the dedispersion DM caused by the intrinsic tilt of the sub-bursts. Main panel: the 234 sub-bursts that were well fit in 2D with error bars. The intrinsic intra-burst drift can cause a burst to be found at a DM that appears to be $\lesssim\SI{12}{\dmu}$ higher than the real DM. Inset: zoom in on bursts with $\sigma_t<\SI{0.6}{\milli\second}$. Considering the linear trend, the difference in the real DM within and between observations is much less than \SI{1}{\dmu}.}
	\label{fig:DMapparent}
\end{figure}

It is very unlikely that the physical reason for the intra-burst drift is a different DM in each burst, because of its tight relationship with the burst width and the small separation in time of many bursts. But given that the intra-burst drift is an intrinsic property, it is still useful to look how it changes the DM that one would measure if one were to observe only a single burst or sub-burst of FRB 121102. To calculate this we can again use Equation~\ref{eq:ddm} with the central frequency $\nu_0$ and the tilt $d_t$ of each burst. The $\delta\DM$ that we obtain is now the apparent DM difference from the DM that was used for dedispersion.

The resulting apparent $\delta\DM$ is shown in Fig.~\ref{fig:DMapparent}. The left panel demonstrates that the intrinsic tilt can make a burst appear to have a DM that is higher by up to \SI{\sim 12}{\dmu}. This illustrates why DMs obtained by maximizing S/N vary a lot. It can even affect structure maximized DMs in the absence of sharp sub-bursts, although differences are likely to be less than \SI{\sim 1}{\dmu}. Note that the values shown here are from sub-bursts, and a structure maximizing code is used for bursts with several sub-bursts. The measured DM therefore depends on the width of each sub-burst, the amplitudes, and how those are weighted within the code.

The inset in Fig.~\ref{fig:DMapparent} allows us to compare differences in the apparent DM between bursts with $\sigma_t<\SI{0.6}{\milli\second}$ between different observations. A change in the real DM between bursts in the same observation would result in a scatter \ed{in $\delta\DM$} independent of $\sigma_t$, whereas a significant change between observations would cause bursts to be offset by that DM. From this consideration, we can put tentative upper limits on these changes of \SI{0.5}{\dmu}.

\subsection{Burst energy distribution}
\label{subsec:energy}

In this section, we present the scaled burst energies as we calculated them in Section~\ref{sec:fitting}.
However, the term burst is not clearly defined, and what we used as bursts and sub-bursts are only practical, empirical differentiations that lack theoretical justification.
We therefore carry out the analysis in this section for three different interpretations of burst energy:
(1) the scaled energy of bursts as they were used in the fitting process, (2) the energy of burst packs as we defined them in Section~\ref{subsec:rates}, where we summed up the energies of bursts that are separated by less than \SI{.1}{\s}, and (3) the energy split between individual sub-bursts according to their volume $V_i$. For all three versions, we use the scaled energies.

We exclude bursts that are close to the edge of the observing band, to ensure that the fit to the spectrum fully includes the burst centres.
This also takes care of biases from narrowband bursts having a higher chance to emit their full energy in the band (given that the burst centre is in the band), and biases from wideband bursts which are more likely to be detected than narrowband bursts if their centre is outside the band.
A smaller observational bias could come from bursts with small bandwidths being less affected by DM smearing, therefore detectable over a larger DM range and more easily classified as real bursts (bursts detected at a single DM are classified as RFI by most pipelines).

The dominating observational effect is that bursts with energies close to the detection limit may or may not be detected depending on other properties, primarily burst width and zenith angle. Modelling these detection effects can be complicated and ultimately requires injections into the search pipeline. The common, simple solution is to exclude bursts below a threshold, above which the majority of bursts are being detected. The completeness threshold of the fluence depends primarily on the burst width, as $\mathcal F_\mathrm{thres}\propto \sigma_t^{1/2}$ (since $\mathcal S_\mathrm{thres}\propto \sigma_t^{-1/2}$). 
We compute the fluence completeness threshold from Equation~\ref{eq:flux} as
\begin{align}
\mathcal{F}_\mathrm{thres} &\approx \frac{(S/N)_\mathrm{thres}\, \mathrm{SEFD}_\mathrm{95\%}\sqrt{\mathrm{FWHM}_{t,95\%}}}{\sqrt{n_\mathrm{p}\,\Delta \nu_\mathrm{band}}} \\
&=\frac{6\times \SI{4.8}{\jansky}\times \sqrt{\SI{12.5}{\milli\second}}}{\sqrt{2\times \SI{580}{\mega\hertz}}}=\SI{0.095}{\jansky\milli\second}\,,
\end{align}
where a subscript $95\%$ denotes the quantile, and $\Delta\nu_\mathrm{band}$ the observing bandwidth. The energy threshold follows from Equation~\ref{eq:energy} and Equation~\ref{eq:Etot} as
\begin{equation}
E_\mathrm{thres} = 4\pi D_L^2  \frac{\mathcal{F}\,\Delta\nu_\mathrm{band}}{1+z}s_{95\%}=\SI{7e37}{erg}\,.
\end{equation}

\begin{figure*}
	\includegraphics[width=\textwidth]{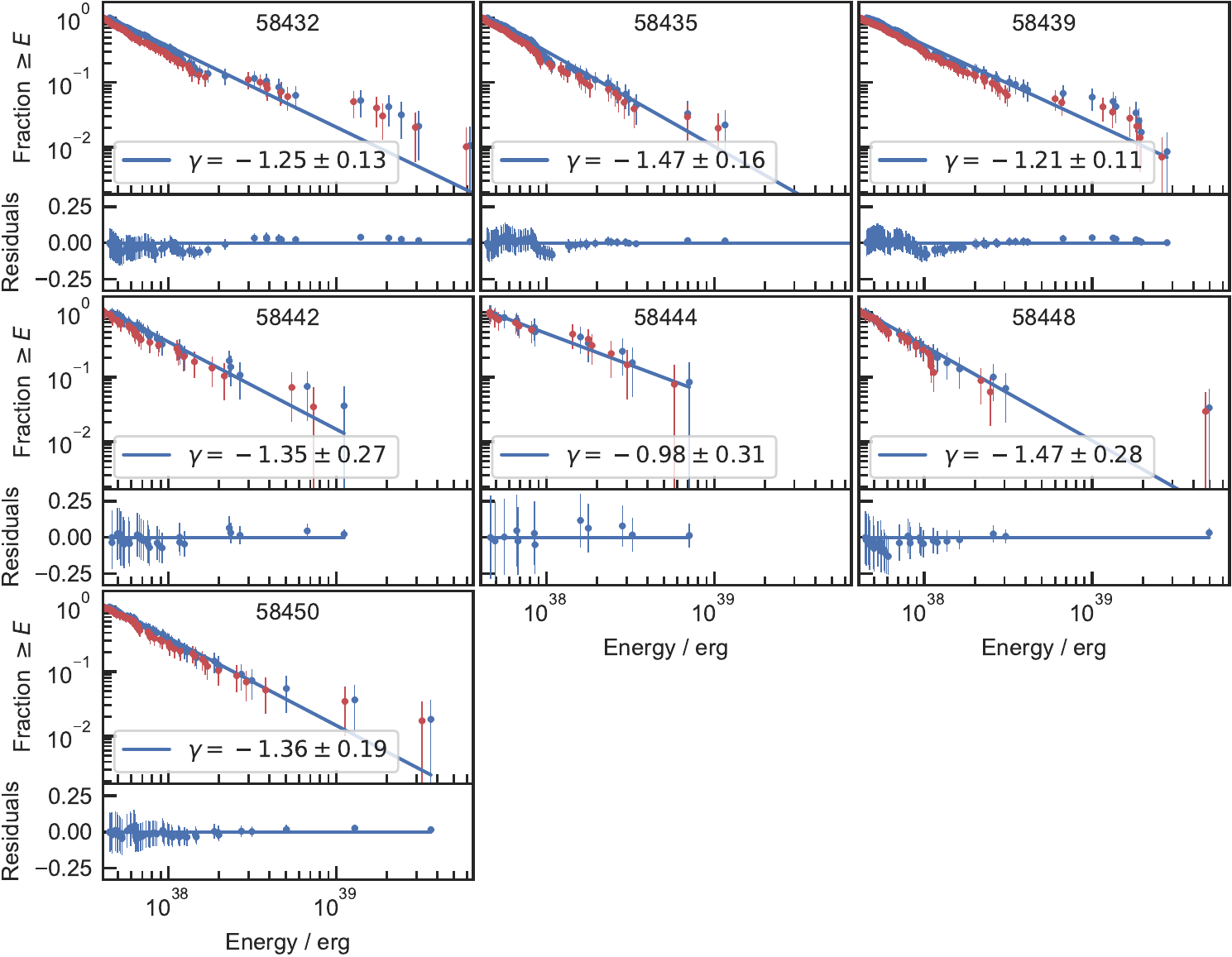}
	\caption{Cumulative burst fraction higher than a given energy, for each observation. Red markers show the measured isotropic-equivalent energies. Blue markers show isotropic-equivalent energies that have been calculated from the measured energy in the band and using the Gaussian fits to extrapolate beyond the band edges. Lower panels show residuals for each observation. Fitted power-law indices are given in the legend. The observation on MJD 58409 is not shown due to insufficient number of bursts.}
	\label{fig:energy_obsi}
\end{figure*}

We show the cumulative energy distribution function above the completeness limit for each observation in Fig.~\ref{fig:energy_obsi}, excluding the observation on MJD 58409, with only two bursts. For comparison, we show the measured in-band energies in red and the scaled energies $E_\mathrm{tot}$ in blue. The errors of the cumulative number $N$ are calculated assuming Poisson statistics via $\sigma(N)=\sqrt{N}$.
We fit a power-law of the form $N(>E)=kE^\gamma$ to the distribution, where $k$ is some constant and $\gamma$ is the power-law index. We use the maximum likelihood method presented in \citet{James2019} \citep[based on][]{Crawford1970}, with the unbiased estimate for $\gamma$ and its uncertainty given by 
\begin{equation}
\frac{1}{\gamma} = - \frac{1}{(N-1)}\sum_i \log \frac{E_i}{E_\mathrm{thres}} \quad \text{and} \quad \sigma(\gamma)=\frac{\gamma}{\sqrt{N-2}}\,.
\end{equation}
The resulting fits are over-plotted in Fig.~\ref{fig:energy_obsi} and the residuals are shown in separate panels. The power-law indices vary between observations, but the uncertainties indicate that the difference is not statistically significant. The later observations from MJD~58442 on-wards seem to be well described by a single power-law, while the three earlier observations (top row) show deviations. The three earlier observations also have the highest burst numbers. Hence, the effect may be subtle and requires a sufficient number of bursts (as opposed to being a time-variable effect). All three show a dip in the burst numbers around an energy of $E\sim\SI{1e38}{erg}$ that is not visible in the later observations.
Observations 58432 and 58439 have significantly more bursts at $E\gtrsim\SI{e39}{erg}$ than one would expect from the power-law fit.

\begin{figure*}
	\begin{subfigure}{.49\textwidth}
		\includegraphics[width=\columnwidth]{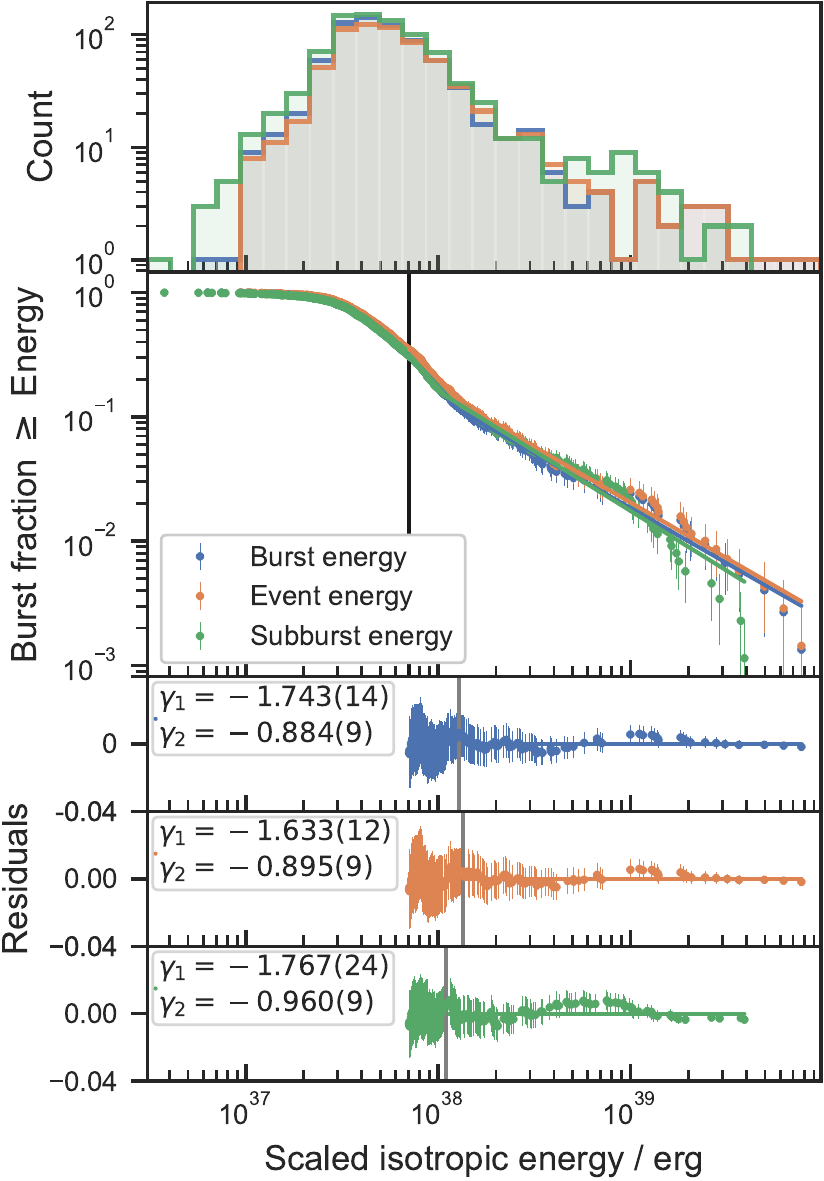}
	\end{subfigure}
	\begin{subfigure}{.49\textwidth}
		\includegraphics[width=\columnwidth]{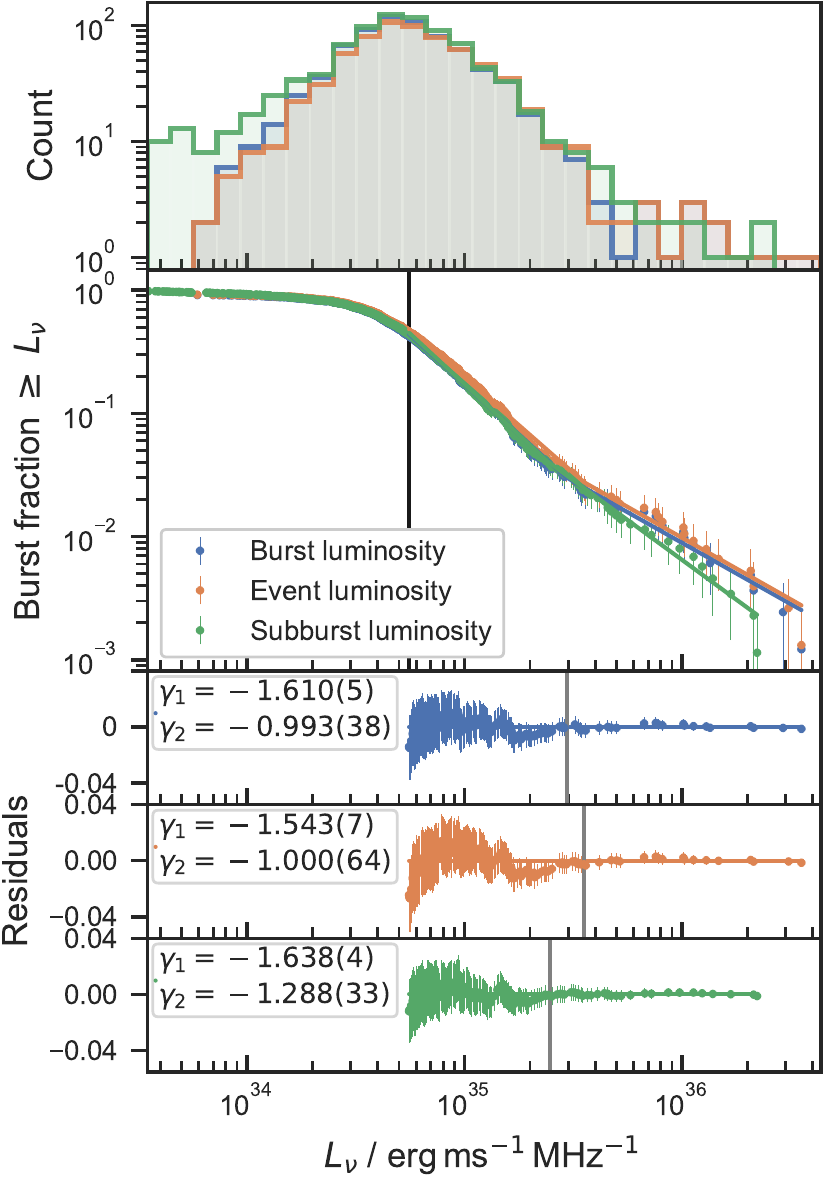}
	\end{subfigure}
	\caption{Cumulative burst fraction higher than a given isotropic-equivalent energy (left) and average specific luminosity (right), summing the bursts from all observations. The differential distribution is shown in the top panel for the three variations. All burst energies (luminosities) are limited to bursts where the burst centres are well within the observing band; and they are scaled up to account for the energy outside the band using the fitted Gaussians. In blue, we plot each burst; in orange, we summed up the energies (luminosities) of bursts that occur within \SI{0.1}{\second}; in green, the energies (luminosities) are distributed among the sub-bursts. Lines mark the broken power-laws that were fit. The three lower panels show the residuals for each fit, with a grey line at the fit break. A black line shows the completeness threshold.}
	\label{fig:energies}
\end{figure*}
\begin{figure*}
	\includegraphics{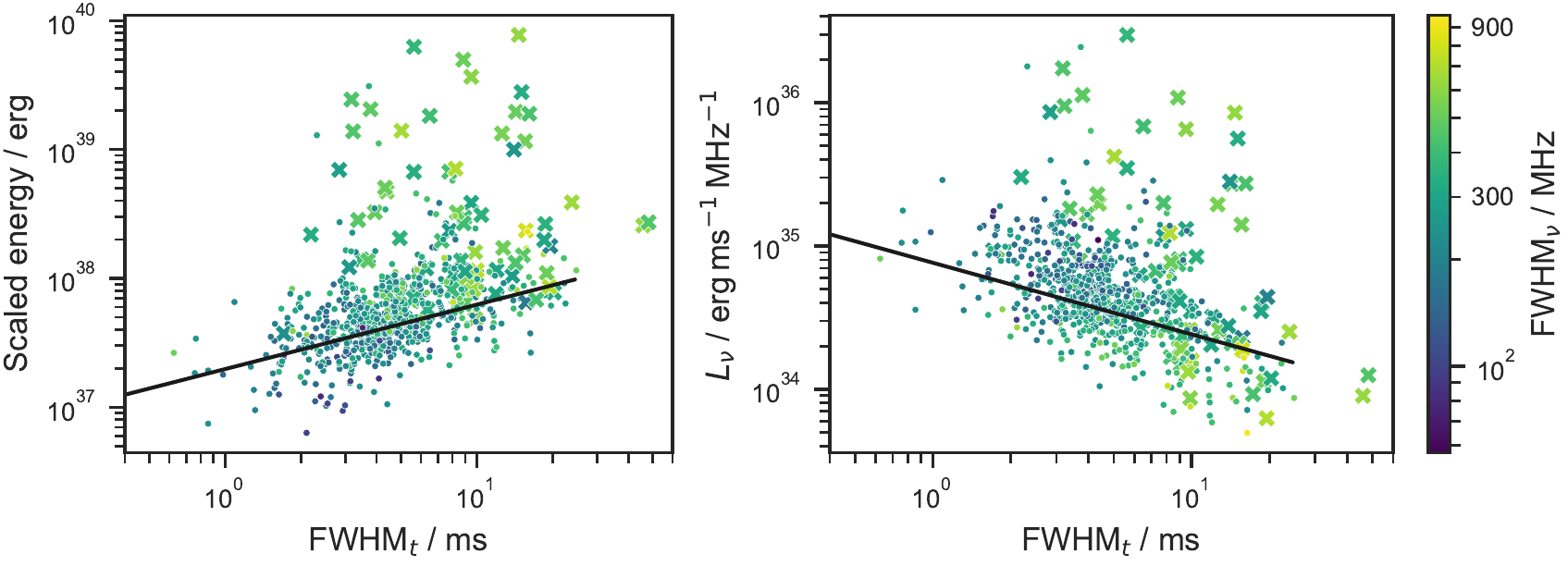}
	\caption{Scaled energies (left) and luminosities (right) plotted against burst duration and bandwidth as colour. Points mark bursts with a single component, while crosses mark bursts with substructure or several sub-bursts. This has the caveat that it is often not possible to identify structure in bursts below a certain $S/N$. A black line marks the completeness threshold.}
	\label{fig:fishtails}
\end{figure*}

We want to explore how the energy distribution depends on the definition of a burst. We show the histogram and the cumulative distribution for three different definitions in the left panel of Fig.~\ref{fig:energies}.
The results show that the different definitions yield similar shapes for the energy distribution. It is so rare (\SI{8}{\percent}) that several bursts occur close to each other that the blue and orange points are almost identical. The incompleteness at energies below the completeness threshold can be seen as a smooth drop in the number density. The sub-burst energies in green do not extend as far to high energies, and the overabundance that was around $E\sim\SI{e39}{erg}$ is shifted to the left. This is because the most energetic bursts all have several sub-bursts among which the energy gets distributed. The dip that was already visible in Fig.~\ref{fig:energy_obsi} is even more pronounced.

It is visible that a single power-law would not fit the cumulative distributions well.
We therefore fit broken power-laws to the three variations that have the form
\begin{equation}
N(>E)=\begin{cases}
k(E/E_\mathrm{break})^{\gamma_1}, & E<E_\mathrm{break}\\
k(E/E_\mathrm{break})^{\gamma_2}, & E\geq E_\mathrm{break}
\end{cases}\,.
\end{equation}
We simultaneously fit the four parameters $\gamma_1$, $\gamma_2$, $k$, and $E_\mathrm{break}$ to minimize the squares of the residuals. The resulting indices are given in Fig.~\ref{fig:energies}. The residuals below $E_\mathrm{break}$ show the dip around $\SI{1e38}{erg}$ that we saw in some observations and a bump below it. They indicate that a power-law might not be a good model for the data below $E_\mathrm{break}$.

Previous studies have placed their focus on the energy distribution; here we want to extend our analysis to the specific luminosity distribution. The isotropic-equivalent specific luminosity can be used as the source intrinsic quantity related to the flux, while the energy is related to the fluence. One reason why the specific luminosity has not received much attention is that the peak flux -- a useful measure for Gaussian shaped bursts -- does not have much value for complex bursts with several peaks. We therefore use the specific isotropic-equivalent luminosity averaged over burst duration and bandwidth, calculated from the scaled energies via $L_\nu=E_\mathrm{tot}/\mathrm{FWHM}_t/\mathrm{FWHM}_\nu$. We calculate the specific luminosity completeness threshold analogously to the energy threshold as
\begin{align}
L_\mathrm{thres} &= 4\pi D_L^2 \frac{(S/N)_\mathrm{thres}\, \mathrm{SEFD}_\mathrm{95\%}\Delta\nu_\mathrm{band}}{\sqrt{n_\mathrm{p}\,\Delta \nu_\mathrm{band}\,\mathrm{FWHM}_{t,5\%}}}\frac{s_{95\%}}{\mathrm{FWHM}_{\nu,\mathrm{median}}}\\
&= 4\pi (\SI{949}{\mega\parsec})^2 \frac{6\times \SI{4.8}{\jansky}\times\SI{580}{\mega\hertz}}{\sqrt{2\times \SI{580}{\mega\hertz}\times\SI{1.9}{\milli\second}}}\,\frac{1.4}{\SI{259}{\MHz}}\\
&=\SI{6.6e31}{erg\per\s\per\Hz}\,,
\end{align}
with the main difference that the 5th percentile instead of the 95th percentile of burst widths has to be used.
The specific luminosity distributions are shown in the right panel of Fig.~\ref{fig:energies}, again for the three variations described above. There is no principle difference in the shapes of the energy and specific luminosity distribution, but some of the features are more pronounced in the luminosity distribution, like the dip around a specific luminosity of \SI{2e35}{erg\per\ms\per\MHz}. The overabundance of bursts at luminosities $>\SI{3e35}{erg\per\ms\per\MHz}$ is less pronounced, resulting in smaller differences between the power-law indices. This difference is smallest when the specific luminosity of sub-bursts (green) is used, this version is therefore most consistent with a single power-law.

To further compare the energy and specific luminosity distributions, we show both plotted against the burst width and bandwidth in Fig.~\ref{fig:fishtails}. Point and cross markers represent single component and multi-component bursts, respectively, as classified by eye. An important caveat is that it is often not possible to identify structure in bursts below a certain $S/N$ \citep[as illustrated in fig.~7 of][]{Gourdji2019}. The completeness threshold can be seen as a black line. Around the energy of \SI{e38}{erg} in the left panel, we can see a steep drop in number density \ed{towards higher energies}. Interestingly, this drop does not seem to be at the same energy for different widths, but rather seems to follow a similar slope as the completeness. Arguably, this is still -- although less -- visible in the right panel.

Complex bursts dominate the distribution at high energies, and the break energies in the broken power-law fits roughly correspond to the energy above which the majority of bursts are complex. It is difficult to tell if this is an observational or a physical effect, because there is a certain energy below which the complex structure of bursts gets buried in noise. This could just coincidentally be similar to the break energy.

We tested if the overabundance of high energy bursts is due to more sub-bursts being identified by looking at the distribution of only the brightest sub-bursts of each burst, but the energy distribution still showed the same features. The larger bandwidth of complex bursts is most likely also an observational effect, with weaker sub-bursts emerging out of the noise.

\subsection{Periodicity}
\label{subsec:period}
The search for short term periodicity focused on the detections on MJD~58435 and MJD~58439, which had the highest detection rate (218 burst per hour) and highest number of detections (227), respectively. Two detection algorithms were applied: the Pearson $\chi^2$ test and the Lomb-Scargle periodogram. 
For both search methods, the range of trial periods searched was \SI{10}{\ms} and \SI{100}{\s}. These values are just below and just above the two peaks in the wait time distribution given in Fig.~\ref{fig:wait_times}. In the case of bursts with multiple sub-bursts, a single arrival time was calculated by taking the weighted mean of the temporal centres of each fitted Gaussian component. 
For the Pearson $\chi^2$ test, the detections were folded with a set of trial logarithmically spaced periods and grouped into eight profile bins. No statistically significant periods were found in either observation. The Lomb-Scargle periodogram was calculated using the \textsc{astropy} LombScargle function. No significant peaks were identified in either observation.

\section{Discussion}
\label{sec:discussion}
\subsection{Rates and wait-times}

Like previous studies of FRB~121102 \citep{Katz2018, Zhang2018a, Gourdji2019,Li2019,Li2021,Aggarwal2021,Hewitt2021} we found a bimodal distribution of wait-times (Fig.~\ref{fig:wait_times}). We showed that the peak on timescales of tens of seconds can be well described by a Poisson process, i.e.\ randomly arriving bursts, with the rate -- that varies between observations -- as the only parameter. We confirmed the results of \citet{Cruces2021} that the two parameter Weibull model is not needed to explain the wait-time distribution within a single observation. Expanding on their conclusion, the Weibull model would also require a varying rate and cannot explain the peak at $<\SI{0.1}{\s}$. It therefore has no advantage over the simpler Poisson model. We also see no advantage in the (two parameter) log-normal distribution that is usually used as an approximation to the Poisson model. Variations in the peak position \citep{Aggarwal2021} contain no information other than the varying rate and different sensitivities of telescopes. \citet{Zhang2018} found a rapidly changing rate in a five-hour observation at 4--\SI{8}{\GHz} with 45 out of 93 bursts arriving in the first 30 minutes. \ed{\citet{Nimmo2022} found a rapidly changing rate for FRB~20200120E at \SI{1.4}{\GHz}.} It remains to be seen if these quick changes can also occur \ed{for FRB~121102} below \SI{4}{GHz}, or if changes are slower, as in the presented observations.

In contrast, the wait time peak at $<\SI{0.1}{\s}$ shows that the bursts sometimes come in packs. Its position is stable over observations with very different rates, and it is therefore reflecting a timescale of the emission process. The wait-times reported by different groups are subject to the different -- sometimes unclear -- distinction between bursts and sub-bursts. The 9 wait-times reported by \citet{Li2019}, which are dominated by data from \citet{Gajjar2018} and \citet{Zhang2018} at frequencies 4--\SI{8}{\GHz}, yield a median of \SI{5.2}{\ms}. \citet{Li2021} report their log-normal fit to peak at \SI{3.4(10)}{\ms}, which is dominated by the wait-times between sub-burst. \citet{Hewitt2021} report \SI{24}{ms}, whereas they did not use wait-times between sub-bursts. The different fitting methods and better sensitivity of FAST -- likely resolving more sub-bursts -- dominate the differences in numbers and can explain the difference between the \SI{3.4(10)}{\ms} and our median of \SI{9.7}{\ms}. Our median of \SI{22}{\ms}, when we exclude wait-times of sub-bursts, is close to the \SI{24}{ms} of \citet{Hewitt2021}, implying no significant change over the \SI{\sim2}{years} between the studied observations. The ambiguity in the definition of sub-bursts also dominates the difference between \citet{Li2021} and the higher value of \SI{39}{\ms} that \citet{Xu2021} find for FRB~20201124A without using wait-times between sub-bursts. Still, compared to our \SI{22}{\ms}, the \SI{39}{\ms} in FRB~20201124A is almost twice as high (at rest frame they translate to \SI{\sim18}{\ms} and \SI{34.5}{\ms} respectively) and show that the peak timescale is different for different sources.

In summary, we see one peak in the wait-times that reflects the occurrence timescale and one that reflects an emission timescale. They have to be interpreted in the light of proposed emission mechanisms, keeping in mind the lack of a strict short-term periodicity. The most popular FRB models involve a neutron star as the central engine of the FRB source \citep{Platts2019,Petroff2021}. Direct observational evidence came from bursts of the magnetar SGR~1935+2154 that would have been observable from an extragalactic distance. They were simultaneously observed in radio \citep{Bochenek2020, CHIME2020} and hard X-rays \citep{Mereghetti2020,Ridnaia2021,Li2021xray,Tavani2021}. Magnetars in X-rays often show several sub-bursts within a rotational phase \citep[see e.g.][]{Huppenkothen2015}, while the arrival of normal pulses follows Poisson statistics \citep{Gogus1999,Gogus2000,Kondratyev2018}. The counterpart in radio to what is seen as sub-bursts in X-rays, can be visible as separate bursts, as was seen in the simultaneous bursts of SGR~1935+2154 \citep[see, e.g., fig.\ 1 in][]{Mereghetti2020}. The clustering in SGR~1935+2154 radio bursts has also been seen by \citet{Kirsten2021}, who detected two bursts in several hundred hours of observations. These two bursts were only \ed{\SI{\sim1.4}{\s}} apart and within the same rotation. Other magnetars \citep{Pearlman2018,Wharton2019} and pulsar giant pulses \citep[see e.g.][]{Karuppusamy2010, Geyer2021} can as well occur multiple times within a rotational period, which produces bimodality in the wait-times, similar to the one we see in FRB~121102.
Apart from these similarities, magnetars emit in radio only in parts of their rotational phase \citep{Pearlman2018}, while no such rotational period has been found in FRB~121102 \citep[and Section~\ref{subsec:period}]{Zhang2018,Cruces2021,Aggarwal2021,Li2021,Hewitt2021}. The Galactic Centre magnetar J1745$-$2900 can, however, emit radio pulses in \SI{\sim70}{\percent} of its rotational phase, although they are clustered in smaller windows \citep{Wharton2019}. In the context of a rotating magnetar with stable emission regions, we can constrain the rotational period to be between the two observed peaks, but several scenarios could make the period unobservable. One possibility is that the emission cone of the source is larger than the angular difference between the rotational axis and our line of sight. This way, bursts would be observable in every part of the rotational phase. Another possibility is that -- unlike in magnetars -- the emission region is not restricted by the magnetic axis and hence the rotational phase.

The stable wait-time peak at $<\SI{0.1}{\s}$ is also in accordance with sub-bursts coming from oscillations in a magnetar crust and core, as proposed by \citet{Wadiasingh2020}. This model is used to explain quasi-periodic oscillations seen in magnetar X-rays that are at similar timescales as sub-bursts of FRB~121102.
In the framework of the synchrotron maser mechanism \citep{Metzger2019,Margalit2020}, the observed emission timescale could be the timescale over which the maser is stable or at least can emit in our observing band in a stable manner.
\citet{Li2019} have argued that the presence of the short wait-time peak would favour models where a neutron star travels through an asteroid belt and FRBs are caused by the collisions between asteroids and the neutron star.
This hypothesis is ruled out by the fact that the $<\SI{0.1}{\s}$ wait-time peak is stable on timescales of years (compare our results here with those of \citet{Hewitt2021}). We can think of no reason for asteroids to consistently cluster such that they would collide with the neutron star on sub-second timescales.

\subsection{General properties}
Earlier observations of FRB 121102 reported in \citet{Gourdji2019} observed a lack of bursts below \SI{1350}{\mega\hertz}, with all 41 bursts being higher than this frequency. \citet{Hewitt2021} confirmed the tendency in a larger set of observations around the same time in September 2016, although they found some bursts below \SI{1350}{\mega\hertz}. We showed that there is no such clear, preferred frequency range in the time window that we reported here.
The change in emission frequency must be due to the difference in time of about 2 years (\num{5} activity cycles). Future observations are needed to investigate if this is a long-term trend to lower frequencies or random variations of the emission band.

\citet{Hessels2019} showed that the DM of bursts is stable on weekly timescales by analysing bright, structured bursts. They also measured the scintillation bandwidth and estimated from it that scattering is negligible at L-band. Nevertheless, some studies have interpreted the sad-trombone effect in unresolved bursts as short term DM variations \citep{Li2021} and intrinsic burst shapes as scattering tails \citep{Aggarwal2021}. We confirmed the conclusions from \citet{Hessels2019} by showing that there is a relationship between burst durations and the intra-burst drift. This relationship must be intrinsic because DM variations would affect all bursts regardless of their temporal width. However, intra-burst drifts by themselves can look like DM variations of several \si{\dmu}. These apparent DM variations do not change the conclusion of other studies on the long-term trend in the DM of FRB~121102 that is of the order of \SI{\sim10}{\dmu} since its discovery \citep{Spitler2014}. It has been investigated by \citet{Hessels2019}, \citet{Oostrum2020}, and \citet{Li2021} using structure optimized DMs, thereby largely mitigating the effect of intrinsic burst drifts. Cosmological applications of FRBs are also not significantly affected by these higher appearing DMs because the uncertainties in the DMs of host galaxies, which are of the order \SI{50}{\dmu}, are much higher. Interestingly, burst C232 has not only sub-bursts that follow the law of the intra-burst drift, but also structure in the last component that is instead straight at the inferred DM. The reason could either be that the drift rate changes in the middle of the burst, or the structure could be a propagation effect that happened shortly after emission.

The temporal width distribution is consistent with the literature on FRB~121102 and shows no long-term temporal variations. In Fig.~\ref{fig:envelopes} complex bursts have systematically larger widths and bandwidths, this is likely a bias and a hint that we are systematically missing additional sub-bursts in weaker bursts.

\subsection{The time-frequency drift}
We have introduced a new way of fitting bursts from repeating FRBs. Previous studies have used various methods. \citet{Hessels2019} used 2D Gaussians (without drift or rotation) to measure the drift of burst centres and compared it to their second method, where a 2D elliptical Gaussian rotated by an angle $\theta$ is fitted it to the 2D-autocorrelation function of bursts. \citet{Rajabi2020} applied the latter technique to sub-bursts to measure the intra-burst drift. \citet{Aggarwal2021} have fitted the bursts dynamic spectra with a more complex function, including a burst-dependent DM and scattering. In this work, we fitted the dynamic spectra with 2D elliptical Gaussians that include a linear shift of the arrival time $t_0$ with frequency. This form has several mathematical and practical advantages. Mathematically it describes the same function as a rotated Gaussian, but a rotation in a space with different dimensions (here time and frequency) makes the parameters lose their physical meaning. On the other hand, $\sigma_t$ and $\sigma_\nu$ in our definition have a clearly defined meaning for all $d_t$. The rotation angle $\theta$ itself is only meaningful as an approximation to our $d_t$.

The sad-trombone drift has previously been measured by several studies \citep{Hessels2019,Josephy2019,Caleb2020} and \citet{Hessels2019} find much stronger drifts than the ones presented here. Only 3 of their 13 analysed bursts, which were observed with the same observing system, have a drift above \SI{-100}{\MHz\per\ms} and the strongest drift is \SI{-286.89(3)}{\MHz\per\ms}. In contrast, the strongest drift in our 12 analysed bursts is \SI{-100.6}{\MHz\per\ms} and our range agrees with the bursts that \citet{Caleb2020} found at similar frequencies, although using a different method and a small sample. The large differences show a significant temporal change in the sad-trombone drifts between September 2016 \citep{Hessels2019} and November 2018.

In Section~\ref{subsec:tilt} we found that within bursts, the intra-burst drift and sad-trombone drift are equal within the uncertainties. Yet, $d_t$ and $\sigma_t$ were seen to be in a \ed{nearly} linear relationship. This excludes the following two simple models. The simplest model would be that in the sad-trombone effect only the frequency centres of sub-bursts drift while sub-bursts are not affected. This would mean only statistical fluctuations of $d_t$ around 0 and no relation between $d_t$ and $\sigma_t$. The second model was described earlier as bursts drifting constantly with drift $d_\nu$ and a Gaussian frequency envelope $w_\nu$ that is superimposed by temporal modulation. For this model, one obtains a quadratic relationship for the measured quantities as $d_t=\sigma_t^2d_\nu/w_\nu^2$. These two simple yet plausible possibilities are in conflict with the \ed{nearly} linear relationship found.

Radius-to-frequency mapping is the common explanation for the sad-trombone drift in models where the emission originates in the magnetosphere of a neutron star \citep[][]{Lyutikov2020b, Tong2022}. A close relation between the sad-trombone drift and the intra-burst drift can be expected in this model, but it does not predict the way in which a burst is modulated in time. The linear $d_t$--$\sigma_t$ therefore does not contradict the radius-to-frequency mapping, but it sets constrains for the way that the temporal structure is created. To our knowledge, there are no predictions for this structure from the emission models.

\ed{The toy model of \citet{Metzger2022} describes the spectro-temporal structure as a Gaussian in frequency, whose central frequency, bandwidth, and flux evolve as power-laws. It can describe the emission in the framework of several physical models like the synchrotron maser model \citep[e.g.][]{Metzger2019}, or radius-to-frequency mapping \citep{Lyutikov2020b}. The toy model can reproduce many properties of FRB~121102 very well, like the sad-trombone effect and shorter bursts with stronger drifts at high frequencies. On the other hand, the model predicts a dependency of duration on bandwidth and frequency, which we do not see in our data. Yet, it might be obscured by the strong $d_t$--$\sigma_t$ correlation. Another feature that the model does not reproduce well is the variety of sub-bursts that we see. Bursts like B29, B180, or C232 show a variety of sub-burst widths and fluxes that require adding individual parameters per sub-burst. Finally, some $d_t$--$\sigma_t$ relationship is expected for the model, but the exact shape has not yet been investigated and requires simulations with realistic parameters.}

So far, the $d_t$--$\sigma_t$ relation is only explained by \citet{Rajabi2020} for a family of models where a relativistically moving FRB source is triggered and only after a delay time $\tau_\mathrm{D}'$ emits radiation at a narrow frequency $\nu_0'$ with a width $\tau_\mathrm{W}'$ in the source's rest frame. The radiation is then observed at a frequency $\nu_\mathrm{obs}$ changed by the relativistic Doppler shift. The delay time and width are equally Doppler shifted and therefore observed as
\begin{equation}
t_\mathrm D=\tau_\mathrm{D}'\frac{\nu_0'}{\nu_\mathrm{obs}}\quad\text{and}\quad
t_\mathrm{W}=\tau_\mathrm{W}'\frac{\nu_0'}{\nu_\mathrm{obs}}\,.
\end{equation}
A slightly faster part of the emission region will be observed with slightly less delay and at a higher frequency with the ratio given by
\begin{equation}
d_t\equiv\dv{t_\mathrm D}{\nu_\mathrm{obs}}=-\tau_\mathrm{D}'\frac{\nu_0'}{\nu_\mathrm{obs}^2}
=-\frac{\tau_\mathrm{D}'}{\tau_\mathrm{W}'}\frac{t_\mathrm{W}}{\nu_\mathrm{obs}} \,.
\end{equation}
The ratio $\tau_\mathrm{D}'/\tau_\mathrm{W}'$ is a constant that depends on the emission mechanism; $t_\mathrm{W}$ is equivalent to our $\sigma_t$. This quick reformulation of the model by \citet{Rajabi2020} showed that the linear $d_t$--$\sigma_t$ is the same that is predicted, and also that the formulation we used is a good description of their model.
Apart from this success, the model does currently not predict a relationship between the sad-trombone drift and the intra-burst drift. It furthermore does not give a reason for the sad-trombone drift to always be negative.

In summary, none of the current models can explain both observational aspects that we discussed. Additional theoretical and observational studies are needed to understand our findings -- and the spectro-temporal structure of FRBs in general -- in the context of the various proposed models. Bursts detected at various frequency bands but within the same activity window could be used to extend our analysis to the relationships between $d_\nu$, $d_t$, $\sigma_\nu$, and $\nu_0$. This has the potential to further test the theory of \citet{Rajabi2020}, the results of \citet{Chamma2021}, \ed{the model of \citet{Tuntsov2021}, as well as the model by \citet{Metzger2019} in different theoretical contexts}.

\subsection{Burst energy distribution}

The high burst rate in our observations allowed us to probe and compare the energy distribution on individual days. We found that a single power-law is an insufficient fit even in single observations due to an excess of high-energy bursts. These bright bursts also tend to have more complex morphologies. It is unclear whether this represents different emission mechanisms or regions, or if this is simply a result of complex morphologies being easier to see in brighter bursts.

Past studies have focused entirely on the analysis of the energy distribution. We showed that instead it is the power or specific luminosity that is dictating the energy distribution, while the distribution of burst width and bandwidth do not vary significantly between low and high energy bursts. However, we found no features in the specific luminosity distribution that were not already visible in the energy distribution.

Past studies of the energy distribution \citep{Gourdji2019, Lin2019,Cruces2021, Aggarwal2021, Hewitt2021} have found different power-law indices, but low numbers and different energy dependent completeness thresholds have complicated the comparisons. \citet{Li2021} found a bimodal energy distribution and that the higher energy bursts were only detectable in their earlier observations so that not only the rate changes but also the form of the energy distribution. We see weak evidence for such a change, e.g.\ comparing observations 58439 and 58448. \ed{The dip in three consecutive observations is around the same energy of \SI{1e38}{erg}, where \citet{Li2021} report a deficiency of bursts.} We also see a similar, yet weaker bimodality in the top panel of Fig.~\ref{fig:energies}. In our data this appears only when the energies of individual sub-bursts are used, but we confirmed with the available data of \citet{Li2021} that the bimodality persists when bursts close in time are summed together. To sum up, we therefore cannot conclusively confirm these important results, but our data suggests that they are not due to a detection bias and that the emission mechanism is indeed time variable. This is also visible in the difference between the energy--width relation of \citet{Hewitt2021} and our data (Fig.~\ref{fig:fishtails}). The clear distinction between high and low energy bursts is not present any more in our data. Furthermore, our value of $\gamma_1=-1.74$ in the broken power-law \ed{in Fig.~\ref{fig:energies}} is much steeper than the $\gamma_1=-1.38$ of \citet{Hewitt2021}. The other values of $\gamma_2=-0.88$ above  $E_\mathrm{break}=\SI{1.28e38}{erg}$ agree roughly with the ones of \citet{Hewitt2021}, which are \ed{$\gamma_2=-1.04$} above $E_\mathrm{break}=\SI{1.15e38}{erg}$. Our values also agree with other previously reported values. The value of \ed{$\gamma=-1.8\pm0.3$} by \citet{Gourdji2019}, which was dominated by low energy bursts, was close to our $\gamma_1$. \citet{Cruces2021} reported \ed{$\gamma=-0.8\pm0.1$} for predominantly high energy bursts, in agreement with our $\gamma_2$. We showed that the uncertainties are dominantly systematic and come from different burst definitions -- i.e.\ the unclear distinction between sub-bursts, bursts, and packs -- and from the estimated completeness threshold.

In the absence of concrete theoretical predictions for the slope of the energy distribution, it is most interesting to compare our findings to the energy distributions of known sources that are related to proposed FRB models.
Normal pulses from pulsars tend to follow a log-normal energy distribution \citep[see e.g.{}][]{Burke-Spolaor2012}, whereas cumulative energies of giant pulses follow a power-law with index $\gamma=\num{-2}$ for the Crab pulsar \citep{Popov2007,Bera2019} and \num{-2.63(2)} for J1823$-$3021A \citep{Knight2007,Abbate2020}. For magnetars a $\gamma$ of \numrange{-0.7}{-.6} was found in X-rays \citep{Gogus1999, Gogus2000}. In radio, the magnetar J1745$-$2900 shows a log-normal distribution with a high energy tail \citep{Lynch2015}. Our value of \num{-0.85} only agrees with the values of magnetar X-ray bursts, but this could also indicate similar underlying statistics rather than a common emission mechanism. For example, this underlying statistics could possibly be described by self-organized criticality \citep[see][for a review]{Aschwanden2016}, as discussed for magnetars by \citet{Huppenkothen2015}.

\section{Conclusions}
\label{sec:conclusion}

With our improved search pipeline, we found 849 bursts in 8 observations during the active period in November 2018. The large number of bursts and the high rate of up to \num{218(16)} bursts per hour allowed us to probe several statistical properties to new precisions and to compare them with burst properties measured at other epochs.
\begin{enumerate}
	\item The new form of Gaussians that was fit to the dynamic spectra showed several advantages, and we recommend future studies to adopt it. On the other hand, \ed{error estimates in} the fitting process could be improved by using Bayesian fits, as was done by \citet{Aggarwal2021}.
	\item The event rates vary strongly between observations of the same active cycle, separated by only a few days.
	\item As in previous studies, a bimodal wait-time distribution is clearly visible.
	We confirm the results of \citet{Cruces2021} with high precision that the peak on timescales of tens of seconds is well fit by Poisson statistics. Therefore, burst arrival times with separation $>\SI{0.1}{\s}$ are best described by a non-stationary Poisson process.
	The peak at $<\SI{0.1}{\s}$ is stable and reflective of a source and emission mechanism timescale. It is consistent with the timescales of magnetar bursts.
	\item The (temporal) intra-burst drift and $\sigma_t$ (the width at $\nu_0$) are related linearly. For 10 out of 12 bursts we find it to be consistent with the sad-trombone drift if not only the burst centres drift but also the emission within sub-bursts. None of the current models can explain both of these findings.
	\item The intra-burst drift is the cause of the apparent short term variations in DM that have been reported. We recommend future studies to use the smallest DM (from the sharpest sub-bursts) to dedisperse bursts at the same DM if the separation is on the order of weeks.
	\item The energy distribution is not well fit by a single power-law, as it shows an overabundance of high energy bursts and a dip around \SI{1e38}{erg} that persist over three consecutive observations. A broken power-law fits the high energy bursts better and yields \ed{$\gamma_1=\num{-1.74(1)}$} below $E_\mathrm{break}=\SI{1.28e38}{erg}$ and \ed{$\gamma_2=\num{-0.88(1)}$} above\ed{. The quoted error includes only statistical uncertainties}.
	\item With the given burst numbers, systematic uncertainties can dominate over statistical uncertainties. The distinction between bursts and sub-bursts influences the location of the wait-time peak at $<\SI{0.1}{\s}$, as well as the energy power-law slope.
	\item The specific luminosity is a more fundamental quantity for FRBs than the energy, but has the disadvantage in complex bursts that the peak luminosity is not well measurable. We encourage further exploration of the burst averaged specific luminosity.
\end{enumerate}

\section*{Acknowledgements}
We thank M.~Cruces for her help with the Weibull fit \ed{and Arun Venkataraman for regularly swapping the hard drives at the Arecibo Observatory according to our needs. We thank Fronefield Crawford for refereeing the manuscript and for his helpful comments, and also Zorawar Wadiasingh and Navin Sridhar for useful hints}.
The Arecibo Observatory is a facility of the National Science Foundation operated under cooperative agreement by the University of Central Florida and in alliance with Universidad Ana G. Mendez, and Yang Enterprises, Inc.
LGS is a Lise Meitner Max Planck independent research group leader and acknowledges funding from the Max Planck Society. 
Research by the AstroFlash group at University of Amsterdam, ASTRON and JIVE is supported in part by an NWO Vici grant (PI Hessels; VI.C.192.045).

%%%%%%%%%%%%%%%%%%%%%%%%%%%%%%%%%%%%%%%%%%%%%%%%%%
\section*{Data Availability}

The results of \ed{the burst fits and some of the derived properties are available as supplementary material. The first ten lines of both tables are in Tables~\ref{tab:bursts} and \ref{tab:properties}. The program \textsc{fix\_gpu\_dropouts} is available at \url{https://github.com/JoschaJ/fix_gpu_dropouts}.} The raw data is available upon reasonable request.



\bibliographystyle{mnras}
\bibliography{novrain} % if your bibtex file is called example.bib

%\begin{thebibliography}{99}
%\end{thebibliography}

%%%%%%%%%%%%%%%%%%%%%%%%%%%%%%%%%%%%%%%%%%%%%%%%%%

%%%%%%%%%%%%%%%%% APPENDICES %%%%%%%%%%%%%%%%%%%%%

\appendix

\section{Different parametrizations for an elliptical Gaussian}
\label{app:gaussians}

Throughout the study, we use two parametrizations for elliptical two-dimensional Gaussians. Here we want to provide additional illustrations to familiarize the reader with the different parameters, in particular the drift rates $d_t$ and $d_\nu$. Further, we will provide the equations to compute one from the other and two additional parametrizations that were not used in the paper.
\begin{figure}
	\includegraphics[width=\columnwidth]{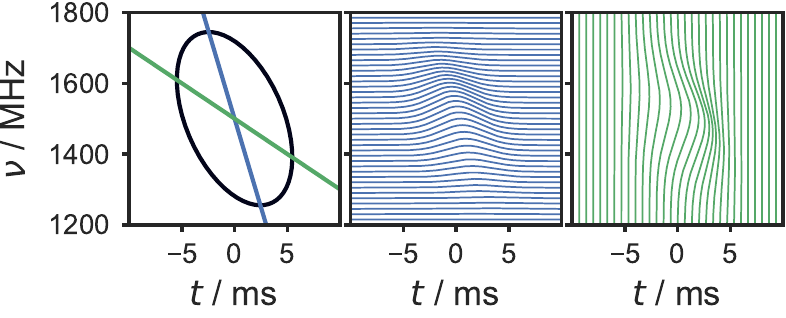}
	\caption{Illustration of the two Gaussian parametrizations. The tree panels show the same elliptical Gaussian. Left panel: a single contour of the Gaussian and the lines along which the Gaussians drift in the temporal and frequency versions, i.e.\ $t_0-d_t(\nu-\nu_0)$ in blue and $\nu_0-d_\nu(t-t_0)$ in green. Middle panel: parametrization that drifts in time ($\mathcal{G}_{\mathrm{2D},t}$), each line can be thought of as one frequency channel. Right panel: parametrization that drifts in frequency ($\mathcal{G}_{\mathrm{2D},\nu}$), each line represents the spectrum at one point in time.}
	\label{fig:gaussians}
\end{figure}

Both are illustrated in Fig.~\ref{fig:gaussians} the properties can be summarized as follows. In the first ($\mathcal{G}_{\mathrm{2D},t}$, Equation~\ref{eq:2Dgaussian}) the central time of arrival $t_0$ drifts linearly in frequency with rate $d_t$ in \si{\ms\per\MHz}, the width $\sigma_t$ is given at $\nu_0$, while $\sigma_\nu$ is the bandwidth of the whole emission, which is equivalent to the bandwidth in a 1D fit to the spectrum. The advantages are that this form is closely related to the DM (see Equation~\ref{eq:ddm} and \ed{the surrounding discussion}), it is therefore also useful for fitting because small DM offsets can be corrected. Lastly, this form is a good description of the FRB model by \citet{Rajabi2020} better yet than the rotated Gaussian in the original description. A disadvantage is that this form is difficult to compare to the sad-trombone drift. 

In the second parametrization ($\mathcal{G}_{\mathrm{2D},\nu}$, Equation~\ref{eq:2DGaussian_nu}) it is the central frequency $\nu_0$ that drifts with rate $d_\nu$ in \si{\MHz\per\ms}. Contrary to the first form, the width $w_t$ is the overall width equivalent to the burst width in the time series, while $w_\nu$ is the bandwidth at $t_0$. The advantage of this form is that it is close to our interpretation of the sad-trombone effect, where we believe that the frequency centres drift with time. A disadvantage in fitting is that the drifts of sharp bursts are strongly susceptible to small DM changes, but this can possibly be used as an advantage in the future to measure the DM with high precision once we fully understand the underlying relations (e.g.\ by not optimizing the structure but instead requiring equality of $d_\nu$ and the sad-trombone drift).

To find the conversion between the two forms, we look at the general form of
an elliptical two-dimensional Gaussian function. It can generally be expressed as 
\begin{equation}
\mathcal{G}(x,y)=A\exp[-\left(a(x-x_0)^2+2b(x-x_0)(y-y_0)+c(y-y_0)^2\right)]\,.
\end{equation}
The Gaussians that we defined in Equations~\ref{eq:2Dgaussian} and \ref{eq:2DGaussian_nu} are also elliptical Gaussians as they can be rewritten in the above form (replacing $x$ by $t$ and $y$ by $\nu$) with
\begin{equation}
a=\frac{1}{2\sigma_t^2}\,,\qquad b=-\frac{d_t}{2\sigma_t^2}\,,\qquad c=\frac{d_t^2}{2\sigma_t^2}+\frac{1}{2\sigma_\nu^2}\,, \label{eq:Gtparams}
\end{equation}
for $\mathcal{G}_{\mathrm{2D},t}$ and
\begin{equation}
a=\frac{1}{2w_t^2}+\frac{d_\nu^2}{2w_\nu^2}\,,\qquad b=-\frac{d_\nu}{2w_\nu^2}\,,\qquad c=\frac{1}{2w_\nu^2}\,,  \label{eq:Gnuparams}
\end{equation}
for $\mathcal{G}_{\mathrm{2D},\nu}$. 
For a given burst the two functions have to be equal and one obtains the conversion relations
\begin{equation}
w_\nu = \frac{\sigma_\nu}{\sqrt{1+\left(\frac{d_t\sigma_\nu}{\sigma_t}\right)^2}}\,,\quad
d_\nu = \frac{d_t}{d_t^2+\frac{\sigma_t^2}{\sigma_\nu^2}}\,,\quad
w_t=\sqrt{\sigma_t^2+(d_t\sigma_\nu)^2}\,.
\label{eq:conversions}
\end{equation}
The reverse relation can simply be obtained from the symmetry by swapping $w\leftrightarrow\sigma$ and $\nu\leftrightarrow t$.

A third form that was not used in this study but is commonly used in statistics and referred to as bivariate normal distribution takes the form
\begin{align}
\mathcal{G}(t,\nu)=A \exp\Bigg[-\frac{1}{(1-\rho^2)}\Bigg(\frac{(t-t_0)^2}{2w_t^2}&-\rho\frac{(t-t_0)(\nu-\nu_0)}{w_t\sigma_\nu}\nonumber \\
&+ \frac{(\nu-\nu_0)^2}{2\sigma_\nu^2}\Bigg)\Bigg]\,,
\label{eq:bivariate}
\end{align}
with $-1<\rho<1$.
It is related to the other forms via
\begin{equation}
\rho=d_t\frac{\sigma_\nu}{w_t},\quad
\rho=d_\nu\frac{w_t}{\sigma_\nu}
\end{equation}
and Equations~\ref{eq:conversions}. $\rho$ describes the tilt of the Gaussian in units of $w_t$ and $\sigma_\nu$, which seems to have no useful physical interpretation in the case of FRBs. However, it might be useful for fitting as the parameters could be less correlated.

A fourth form is the Gaussian that is rotated by an angle $\theta$ and was used in earlier studies. Its parameters have no physical meaning, and we recommend using one of the other three forms instead. \ed{Results from previous studies that used the rotated form can be converted} as follows.
The rotated Gaussian is given with respect to the general form by
\begin{align}
a&=\frac{\cos^2\theta}{2\sigma_x^2} + \frac{\sin^2\theta}{2\sigma_y^2}\,,\\
2b&=-\frac{\sin(2\theta)}{2\sigma_x^2} + \frac{\sin(2\theta)}{2\sigma_y^2}\,,\qq{and}\\
c&=\frac{\sin^2\theta}{2\sigma_x^2} + \frac{\cos^2\theta}{2\sigma_y^2}\,.
\end{align}
Note how these equations are already in conflict with $\sigma_x$ and $\sigma_y$ having different units. Ignoring these violations of mathematical rules this form can be set equal to Equation~\ref{eq:bivariate} and after some time one can obtain
\begin{align}
\rho&=\frac{k\sin(2\theta)/2}{\sqrt{1+(k\sin(2\theta)/2)^2}}\,,\qq{with}
k=\frac{\sigma_x}{\sigma_y}-\frac{\sigma_y}{\sigma_x}\,,\\
w_t&=\sigma_x\sigma_y\sqrt{\frac{1+(k\sin(2\theta)/2)^2}{\sigma_x^2\sin^2\theta+\sigma_y^2\cos^2\theta}}\,,\\
\sigma_\nu&=\sigma_x\sigma_y\sqrt{\frac{1+(k\sin(2\theta)/2)^2}{\sigma_x^2\cos^2\theta+\sigma_y^2\sin^2\theta}}\,,
\end{align}
or reversely
\begin{align}
\tan(2\theta) &= \frac{2\rho}{w_t\sigma_\nu(\sigma_\nu^{-2}-w_t^{-2})}\qq{and}\\
\sigma_x^2&=\frac{2(1-\rho^2)w_t\sigma_\nu}{\frac{\sigma_\nu}{w_t}+\frac{w_t}{\sigma_\nu}+\mathrm{sgn}(w_t^{-2}-\sigma_\nu^{-2})(w_t^2\sigma_\nu^{-2}+w_t^{-2}\sigma_\nu^2-2+4\rho^2)^{1/2}}\,.
%d_t=\frac{\cos2\theta-\sqrt{1-\frac{\sigma_t^2}{\sigma_\nu^2}\sin^2 2\theta}}{\sin2\theta}
\end{align}

\section{Derivation of $\delta\DM$}
\label{subsec:taylorDM}

Doing the full Taylor expansion of $\Delta t$ yields
\begin{align}
\Delta t&=\sum_{n=0}^{\infty}\eval{\frac{\Delta t^{(n)}(\nu)}{n!}}_{\nu=\nu_0}(\nu-\nu_0)^n\\
&=\eval{\Delta t(\nu_0)+\dv{\Delta t}{\nu}}_{\nu=\nu_0}(\nu-\nu_0)+\order{\nu^2}  \\
&=a\times\delta\DM\left(\frac{1}{\nu_0^2}-\frac{1}{\nu_\mathrm{ref}^2}\right)-2a\times\delta\DM\frac{1}{\nu_0^3}(\nu-\nu_0)+\order{\nu^2}\,.
\end{align}
Dropping the first term as we only care about the shift with respect to $\nu_0$ we get
\begin{equation}
\Delta t\approx-2a\times\delta\DM\frac{1}{\nu_0^3}(\nu-\nu_0)
\end{equation}

\section{Supplementary material}
\begin{figure}
	\begin{subfigure}{.49\textwidth}
		\includegraphics[width=\columnwidth]{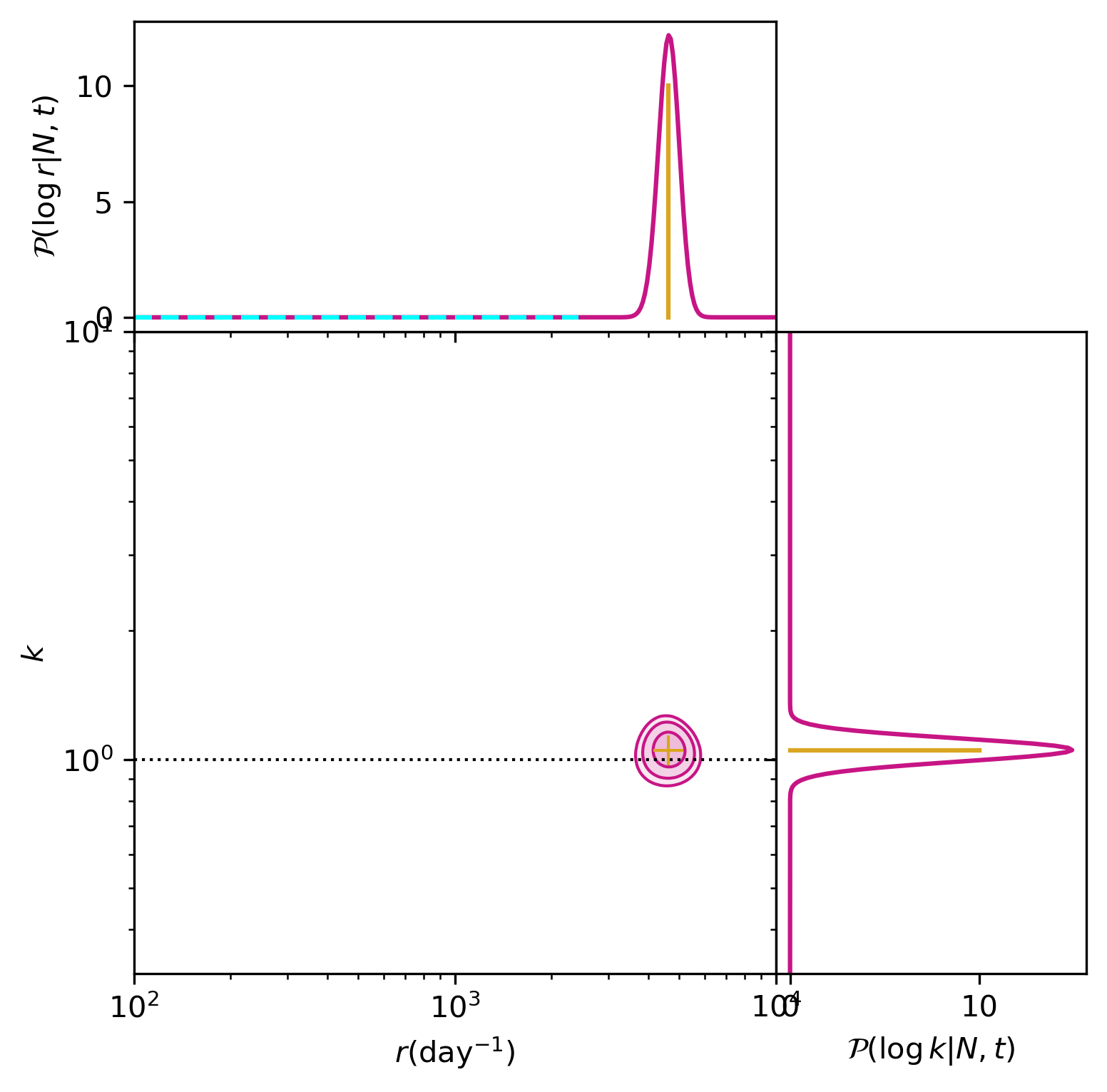}
	\end{subfigure}
	\begin{subfigure}{.49\textwidth}
		\includegraphics[width=\columnwidth]{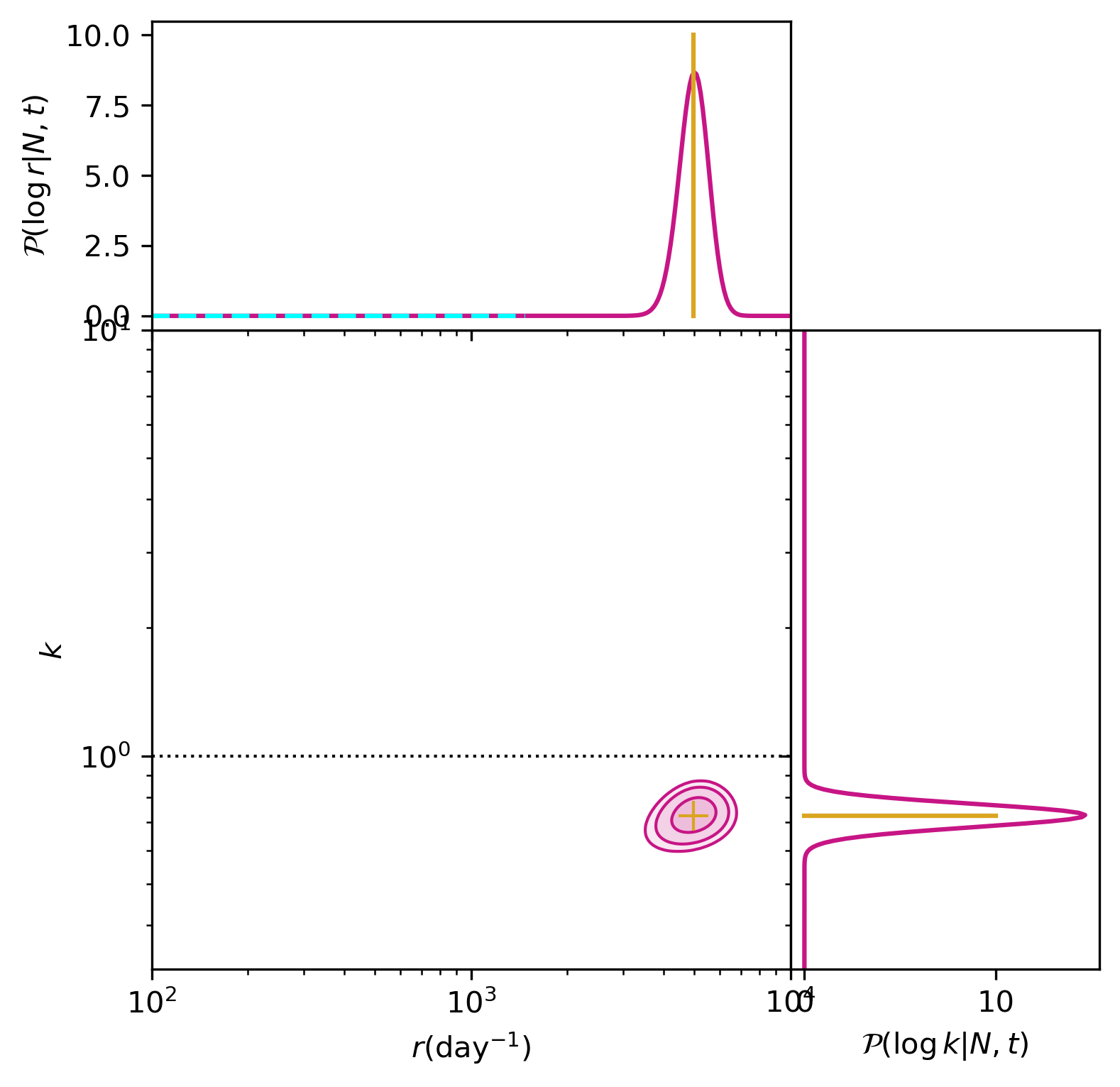}
	\end{subfigure}
	\caption{Posterior distribution for the Weibull parameters in the observation on MJD 58435 \ed{for all bursts (bottom plot) and for burst packs where we averaged over the arrival times of bursts that arrived within $\delta<\SI{0.1}{\s}$ (top plot). In the latter case where wait times $\delta<\SI{0.1}{\s}$ are excluded the arrival times are consistent with a Poisson distribution ($k=1$).}}
	\label{fig:weibull}
\end{figure}

\begin{table*}
	\caption{The first 10 lines of the data table that is available \ed{as supplementary material}. The times of arrival are barycentre corrected and dispersion corrected to infinite frequency using the dispersion constant \ed{$a=\SI[separate-uncertainty=false]{4.1488064239(11)}{\GHz\squared \cm\cubed \per\parsec \ms}$} \citep{Kulkarni2020}.}
	\label{tab:bursts}
	\sisetup{
		separate-uncertainty = false,
		%table-alignment-mode = format,
		%table-number-alignment = center,
	}
	
	\begin{tabular}{@{}ll
			S[table-format = 5.10]
			S[table-format = 2.9]
			S[table-format = -1.7]
			S[table-format = 4.7]
			S[table-format = 1.5]
			S[table-format = 4.5]
			S[table-format = 3.4]
			S[table-format = 1.2e2]%, round-mode = places, round-precision = 2
			@{}}
		\hline
		\hline
		{ID} & {sub-burst} & {TOA / MJD} & {$A$} & {$d$ / \si{\ms\per\MHz}} & {$t_0$ / s} & {$\sigma_t$ / ms} & {$\nu_0$ / MHz} & {$\sigma_\nu$ / MHz} & {Scaled Energy / erg}\\
		\hline
		A1	&	sb1	&	58409.3564972816	&	0.64	(	30	) & 	-0.0048	(	34	) & 	548.9640	(	4	) & 	0.66	(	30	) & 	1430	(	63	) & 	132	(	65	) & 	7.76e+37	\\
		A2	&	sb1	&	58409.3643956201	&	0.56	(	38	) & 	-0.0053	(	82	) & 	1231.3234	(	11	) & 	0.63	(	44	) & 	1700	(	185	) & 	107	(	151	) & 	3.90e+37	\\
		A3	&	sb1	&	58409.3708097451	&	0.15	(	31	) & 	-0.0044	(	103	) & 	1785.4575	(	21	) & 	0.52	(	104	) & 	1577	(	416	) & 	162	(	470	) & 	2.37e+37	\\
		A4	&	sb1	&	58409.3708099014	&	0.17	(	28	) & 	0.0125	(	465	) & 	1785.4710	(	38	) & 	1.26	(	212	) & 	1687	(	244	) & 	82	(	235	) & 	2.55e+37	\\
		A5	&	sb1	&	58409.3765544552	&	0.63	(	80	) & 	-0.0007	(	90	) & 	2281.7590	(	10	) & 	0.64	(	68	) & 	1741	(	548	) & 	146	(	344	) & 	8.56e+35	\\
		A5	&	sb2	&	58409.3765544853	&	0.54	(	102	) & 	-0.0008	(	97	) & 	2281.7616	(	18	) & 	0.86	(	120	) & 	1778	(	961	) & 	200	(	520	) & 	1.37e+36	\\
		A5	&	sb3	&	58409.3765544864	&	82.55	(	1558104	) & 	-0.0016	(	76	) & 	2281.7617	(	846	) & 	0.47	(	64	) & 	3061	(	52502	) & 	436	(	7972	) & 	2.47e+38	\\
		A5	&	sb4	&	58409.3765545378	&	1.06	(	161	) & 	-0.0045	(	37	) & 	2281.7662	(	39	) & 	0.76	(	35	) & 	1828	(	845	) & 	250	(	411	) & 	2.96e+36	\\
		B6	&	sb1	&	58432.2646080422	&	0.14	(	40	) & 	0.0074	(	676	) & 	13.7812	(	114	) & 	1.71	(	506	) & 	1187	(	1319	) & 	148	(	960	) & 	4.49e+37	\\
		B6	&	sb2	&	58432.2646081021	&	0.18	(	57	) & 	0.0015	(	145	) & 	13.7863	(	21	) & 	0.56	(	191	) & 	1317	(	764	) & 	183	(	805	) & 	2.39e+37	\\
		\hline
	\end{tabular}
\end{table*}

\begin{table*}
	\raggedright
	\sisetup{
		separate-uncertainty = false,
		%table-alignment-mode = format,
		table-number-alignment = center,
	}
	\begin{tabular}{@{}
			S[table-format = 5.10]
			S[table-format = 1.5]
			S[table-format = 4.7]
			S[table-format = 1.6]
			S
			S
			@{}}
		\hline
		\hline
		{TOA$_\mathrm{1D}$ / MJD} & $A_\mathrm{1D}$ & {$t_{0,\mathrm{1D}}$ / s} & {$\sigma_{t,\mathrm{1D}}$ / ms} & {$\nu_{0,\mathrm{1D}}$ / MHz} & {$\sigma_{\nu,\mathrm{1D}}$ / MHz}\\
		\hline
		58409.356497281	&	4.54	(	27	) & 	548.96394	(	6	) & 	0.912	(	64	) & 	1429	(	7	) & 	122	(	7	)	\\
		58409.3643956236	&	2.76	(	29	) & 	1231.32370	(	9	) & 	0.752	(	90	) & 	1688	(	14	) & 	90	(	13	)	\\
		58409.3708097508	&	0.96	(	27	) & 	1785.45799	(	39	) & 	1.217	(	394	) & 	1972	(	964	) & 	369	(	409	)	\\
		58409.3708099035	&	1.10	(	21	) & 	1785.47119	(	33	) & 	1.543	(	335	) & 	2237	(	2297	) & 	477	(	774	)	\\
		58409.3765544568	&	3.99	(	33	) & 	2281.75916	(	7	) & 	0.637	(	75	) & 	1763	(	52	) & 	158	(	29	)	\\
		58409.3765544868	&	3.82	(	32	) & 	2281.76175	(	9	) & 	0.809	(	127	) & 	1863	(	128	) & 	229	(	55	)	\\
		58409.3765545116	&	3.42	(	38	) & 	2281.76389	(	8	) & 	0.501	(	84	) & 				&					\\
		58409.3765545488	&	6.07	(	26	) & 	2281.76711	(	5	) & 	1.020	(	54	) & 	1741	(	38	) & 	215	(	23	)	\\
		58432.2646080413	&				&				&				&	1059	(	308	) & 	227	(	142	)	\\
		58432.2646081035	&				&				&				&	1310	(	16	) & 	157	(	18	)	\\
		\hline
	\end{tabular}
\end{table*}

\begin{table*}
	\raggedright
	\begin{tabular}{@{}lllSSS@{}}
		\hline
		\hline
		{Class} & {Diffuse/Tail} & {Dropouts} & {Fluence / Jy ms} & {Downsampling} & {$t_\mathrm{fit}$ / ms} \\
		\hline
		default	&	False	&	False	&	0.1412	&	8	&	5	\\
		default	&	False	&	False	&	0.0709	&	8	&	5	\\
		default	&	False	&		&	0.0432	&	8	&	5	\\
		default	&	False	&	False	&	0.0465	&	8	&	5	\\
		multi	&	False	&	False	&	0.4578	&	8	&	5	\\
		multi	&	False	&	False	&	0.4578	&	8	&	5	\\
		multi	&	False	&	False	&	0.4578	&	8	&	5	\\
		multi	&	False	&	False	&	0.4578	&	8	&	5	\\
		multi	&	False	&	False	&	0.6717	&	8	&	5	\\
		multi	&	False	&	False	&	0.6717	&	8	&	5	\\
		\hline
	\end{tabular}
\end{table*}

\begin{table*}
	\caption{\ed{Some of the burst properties derived throughout the paper. The full table is available as supplementary material. The TOA is averaged over sub-burst TOAs. Time and frequency envelopes as well as $\nu_\mathrm{cent}$ are explained in Section~\ref{subsec:properties} and shown in Fig.~\ref{fig:envelopes}, values are missing if the fits were not successful. The fluence was measured using a frequency and zenith angle dependent SEFD, as described in Section~\ref{sec:fitting}. The scaled energy is the isotropic equivalent energy that is scaled from the fluence and the 2D Gaussian fits via Equation~\ref{eq:Etot}, it is only present for bursts where the 2D fits had reasonable uncertainties. The average specific luminosity is derived in Section~\ref{subsec:energy}.}}
	\label{tab:properties}
	%\raggedright
	\begin{tabular}{@{}ll
			S[table-format = 5.10]
			S[table-format = 2.2]
			S[table-format = 3.0]
			S[table-format = 4.0]
			S[table-format = 1.3]
			S[table-format = 1.2e2]
			S[table-format = 1.2e2]
			r@{}}
		\hline
		\hline
		{Observation}	&	{ID}	&	{TOA}	&	{Time envelope}	&	{Frequency envelope}	&	{$\nu_\mathrm{cent}$}	&	{Fluence}	&	{Scaled energy}	&	{Specific luminosity}	&	{Search ID}\\
		&	&	{(MJD)} & {(ms)} & {(MHz)} & {(MHz)} & {(\si{\jansky\ms})} & {(erg)} & {(\si{erg\per\ms\per\MHz})} & \\
		\hline
		58409	&	A1	&	58409.3564972816	&	2.13	&	310	&	1430	&	0.141	&	7.58E+37	&	1.37E+35	&	2798	\\
		58409	&	A2	&	58409.3643956201	&	1.77	&	213	&	1688	&	0.071	&	4.93E+37	&	1.56E+35	&	2857	\\
		58409	&	A3	&	58409.3708097451	&		&		&		&	0.043	&		&		&	3995-1	\\
		58409	&	A4	&	58409.3708099014	&		&		&		&	0.046	&		&		&	3995	\\
		58409	&	A5	&	58409.3765545122	&	9.90	&	644	&	1789	&	0.458	&		&		&	57	\\
		58432	&	B6	&	58432.2646081847	&	9.51	&	309	&	1304	&	0.672	&	3.86E+38	&	1.56E+35	&	5095	\\
		58432	&	B7	&	58432.2647338737	&	2.56	&	182	&	1704	&	0.043	&		&		&	7732	\\
		58432	&	B8	&	58432.2650200047	&	3.12	&	272	&	1417	&	0.050	&	2.64E+37	&	3.72E+34	&	8343	\\
		58432	&	B9	&	58432.2652493147	&	18.97	&	517	&	1500	&	0.198	&	1.10E+38	&	1.34E+34	&	7500	\\
		58432	&	B10	&	58432.2652784695	&	2.11	&	636	&	1448	&	0.051	&	3.63E+37	&	3.22E+34	&	7963	\\
		\hline
	\end{tabular}
\end{table*}
%%%%%%%%%%%%%%%%%%%%%%%%%%%%%%%%%%%%%%%%%%%%%%%%%%

% Don't change these lines
\bsp  % typesetting comment
\label{lastpage}
\end{document}